\documentclass[prd,aps,showpacs,epsf,floats,onecolumn]{revtex4}%
\usepackage{amssymb}
\usepackage{amsfonts}
\usepackage{amsmath}
\usepackage{graphicx}%
\providecommand{\U}[1]{\protect\rule{.1in}{.1in}}
\begin{document}
\title{\textbf{Complexity of Quantum-Mechanical Evolutions from Probability
Amplitudes}}
\author{\textbf{Carlo Cafaro}$^{1,2}$, \textbf{Leonardo Rossetti}$^{3,1}$,
\textbf{Paul M.\ Alsing}$^{1}$}
\affiliation{$^{1}$University at Albany-SUNY, Albany, NY 12222, USA}
\affiliation{$^{2}$SUNY Polytechnic Institute, Utica, NY 13502, USA}
\affiliation{$^{3}$University of Camerino, I-62032 Camerino, Italy}

\begin{abstract}
We study the complexity of both time-optimal and time sub-optimal quantum
Hamiltonian evolutions connecting arbitrary source and a target states on the
Bloch sphere equipped with the Fubini-Study metric. This investigation is
performed in a number of steps. First, we describe each unitary
Schr\"{o}dinger quantum evolution by means of the path length, the geodesic
efficiency, the speed efficiency, and the curvature coefficient of its
corresponding dynamical trajectory linking the source state to the target
state. Second, starting from a classical probabilistic setting where the
so-called information geometric complexity can be employed to describe the
complexity of entropic motion on curved statistical manifolds underlying the
physics of systems when only partial knowledge about them is available, we
transition into a deterministic quantum setting. In this context, after
proposing a definition of the complexity of a quantum evolution, we present a
notion of quantum complexity length scale. In particular, we discuss the
physical significance of both quantities in terms of the accessed (i.e.,
partial) and accessible (i.e., total) parametric volumes of the regions on the
Bloch sphere that specify the quantum mechanical evolution from the source to
the target states. Third, after calculating the complexity measure and the
complexity length scale for each one of the two quantum evolutions, we compare
the behavior of our measures with that of the path length, the geodesic
efficiency, the speed efficiency, and the curvature coefficient. We find that,
in general, efficient quantum evolutions are less complex than inefficient
evolutions. However, we also observe that complexity is more than length.
Indeed, longer paths that are sufficiently bent can exhibit a behavior that is
less complex than that of shorter paths with a smaller curvature coefficient.

\end{abstract}

\maketitle

\section{Introduction}

Quantum complexity in theoretical physics is a highly interesting quantity.
For example, regarding black holes as holographically described by a
collection of qubits that evolve under Hamiltonians that exhibit fingerprints
of maximal quantum chaos, complexity can be used to specify the phenomena of
particle growth and information spreading near black hole horizons
\cite{susskind16}. In this context, the so-called holographic complexity
(i.e., the complexity of a holographic quantum state) is related to the growth
of the volume in the bulk geometry beyond the event horizon in black holes
\cite{stanford14,leo16B,leo16,belin22}. In many-body physics, complexity can
be used to distinguish highly entangled quantum states of condensed matter
systems and to characterize thermalization and transport in quantum many-body
systems \cite{sondhi21,ali19,ali20,vijay20}. The usefulness of complexity, in
this context, emerges once the link between quantum chaos and complexity in
chaotic quantum systems is made transparent \cite{ali20,vijay20,gomez20}.

There are several notions of complexity in quantum physics. For instance, one
can speak of the complexity of a quantum state (state complexity,
\cite{chapman18,iaconis21}), the complexity of a quantum circuit (circuit
complexity, \cite{nielsen,fernando21}) or, alternatively, the complexity of an
operator (operator complexity, \cite{parker19,roy23,liu23,caputa22}). Despite
their differences, like most measures of complexity, each one of these
measures captures the essential idea that the complexity of a composite object
scales like the number of elementary elements needed to construct it
\cite{lof77,ay08,vijay22}. Depending on the specific context, the scaling with
the number of elementary elements can be conveniently expressed in terms of
geometrically intuitive concepts like lengths and volumes. In the context of
theoretical computer science, Kolmogorov suggested in Ref. \cite{kolmogorov68}
that the complexity of a sequence can be expressed as the length of the
shortest Turing machine program that generates it. In information theory,
Rissanen proposed in Refs. \cite{rissanen78,rissanen86} that the minimal code
length of an ensemble of messages, averaged over the ensemble, specifies the
complexity of an ensemble of messages in information theory.

The relevance of the concepts of length and volumes is essential in the
characterization of complexity in a quantum mechanical setting as well. For
example, the state complexity is the complexity of a quantum state described
in terms of the smallest local unitary circuit that can generate the state
from an initial simple (i.e., factorizable) reference quantum state.
Interestingly, a geometric formulation of the state complexity was first
proposed while investigating the complexity of quantum states in continuous
many-body systems. Specifically, the state complexity of a target state
generated by acting on a source state with a set of parametrized unitary
operators was expressed as the length of the shortest path measured with the
Fubini-Study metric and associated with an allowed realization of the unitary
operator \cite{chapman18}. Moreover, quantum circuits are made of quantum
gates that act on quantum states. In particular, the circuit complexity
specifies the number of primitive gates in the smallest circuit that
implement\textbf{ }a transformation that acts on a given quantum state
\cite{nielsen,fernando21}. It is discrete in nature and, in addition, is a
natural complexity measure for scientists trying to actually build quantum
circuits out of component gates. Interestingly, the definitions of state and
circuit complexities are not unrelated since the state complexity is the
complexity of the least complex unitary operator that connects source and the
target states. The geometric characterization of the concept of circuit
complexity was proposed by Nielsen and collaborators in Refs.
\cite{mike06,mike06B,mike08}. In this geometric formulation, the circuit
complexity of a unitary operator $U$ is continuous in nature and is equal to
the length of the minimal geodesic on the unitary group that joins the
identity operator to $U$. The length of these geodesic curves, in turn,
provide a lower bound for the minimum number of quantum gates needed to
generate the unitary operator $U$. Finally, operator complexity describes the
temporal growth of the size of an operator under the Heisenberg or Lindblad
time evolutions for closed and open quantum systems, respectively. When
evaluated with respect to the so-called Krylov basis, the operator complexity
is called Krylov complexity \cite{parker19}. For a presentation of useful
techniques to handle Nielsen's complexity of nontrivial dynamical systems, we
refer to Refs. \cite{craps22,craps24B}. Furthermore, for a description of a
rather surprising connection between Krylov's and Nielsen's complexities, we
suggest Ref. \cite{craps24}. In Ref. \cite{caputa22}, while studying the
temporal change of the displacement operator during the unitary evolution of
many-body quantum systems governed by symmetries, the expectation value of the
Krylov complexity operator was shown to be equal to the volume of the
corresponding classical phase space. The link between Krylov complexity and
volume was further discussed in the context of specific quantum field
theoretic settings in Ref. \cite{roy23}. In particular, the scaling of this
complexity with the volume was justified by demonstrating that the Krylov
complexity is equal to the average particle number. Then, applying the
proportionality between the volume and the average particle number, the link
was properly motivated.

In Ref. \cite{carloPRD}, we presented an information geometric theoretical
approach to describe and, to a certain extent, comprehend the complex behavior
of evolutions of quantum systems in pure and mixed states. The comparative
analysis was probabilistic in nature, it employed a complexity measure that
relied on a temporal averaging procedure along with a long-time limit, and was
limited to analyzing expected geodesic evolutions on the underlying manifolds.
In particular, we analyzed the complexity of geodesic paths on the manifolds
of single-qubit pure and mixed quantum states equipped with the Fubini-Study
metric \cite{braunstein94} and the Sj\"{o}qvist \cite{erik20} metric,
respectively. We demonstrated in an analytical fashion that the evolution of
mixed quantum states in the Bloch ball is more\ complex than the evolution of
pure states on the Bloch sphere. We also confirmed that the ranking based on
our proposed measure of complexity, a quantity that specifies the asymptotic
temporal behavior of an averaged volume of the region explored on the manifold
during the evolution of the systems, concurs with the ranking based on the
geodesic length of the corresponding paths. Last but not least, we found that
the complexity of the Bures manifold was softened compared to the Sj\"{o}qvist
manifold when we focused on geodesic lengths and curvature properties
\cite{carloPRA23}.

In this paper, going beyond the work of Ref. \cite{carloPRD} and given the
fundamental role played by the concepts of length and volume in the
characterization of the complexity of quantum processes, the goal is to
provide a clear comparative analysis of the complexity of quantum evolutions,
both geodesic and nongeodesic, on the Bloch sphere for two-level quantum
systems. In particular, to enhance our comprehension of the physical
significance of the newly proposed concept of complexity, the comparative
analysis is carried out also in terms of additional quantifiers with an
established physical meaning. The list of these additional measure includes
path lengths, geodesic and speed efficiencies and, finally, curvature
coefficients of the quantum evolutions.

To the best of our knowledge, we aim to tackle a range of unexplored questions
in the literature, some examples of which include:

\begin{enumerate}
\item[{[i]}] Does the complexity of a quantum evolution from an initial to a
final state under a given Hamiltonian encode information beyond the one
encoded in the concept of length of the quantum evolution path?

\item[{[ii]}] Are complex quantum evolutions necessarily characterized by a
large amount of wasted energy?

\item[{[iii]}] Are quantum evolutions characterized by longer paths
necessarily more complex than the ones specified by shorter paths connecting
the same initial and final quantum states?

\item[{[iv]}] Is time-optimality a synonymous with reduced complexity?

\item[{[v]}] Can quantum evolutions that occur at lower quantum speeds and
with larger waste of energy resources be less complex than very fast
evolutions that happen with negligible amounts of wasted energy?
\end{enumerate}

The rest of the paper is organized as follows. In Section II, we introduce
optimal and sub-optimal quantum Hamiltonian evolutions connecting arbitrary
source (i.e., initial) and a target (i.e., final) states on the Bloch sphere
equipped with the Fubini-Study metric. In particular, we characterize both
evolutions in terms of the concepts of path length
\cite{wootters81,provost80,cafaro23}, geodesic efficiency
\cite{anandan90,cafaro20}, speed efficiency \cite{uzdin12}, and curvature
coefficients \cite{alsing24A,alsing24B} of the dynamical trajectories that
connect the given initial and final states. In Section III, we begin by
considering a classical probabilistic setting where the so-called information
geometric complexity is generally used to characterize the complexity of
entropic motion on curved statistical manifolds underlying the physics of
systems when only limited information about them is available
\cite{cafaro07,cafarothesis,cafaro17,cafaro18}. Then, moving to a
deterministic quantum setting, we present our definition of complexity of a
quantum evolution along with a notion of quantum complexity length scale. We
explicitly discuss the physical meaning of both concepts and, in particular,
we discuss the fact that we express both concepts in terms of accessed and
accessible parametric volumes\ of the regions on the Bloch sphere that specify
the evolution from the source to the target states. In Section IV, we
calculate the complexity measure and the complexity length scale of each one
of the two (families of) quantum evolutions mentioned in Section II. In
particular, for each quantum evolution, we compare the complexity measure and
the complexity length scale with the path length, the geodesic efficiency, the
speed efficiency, and the curvature coefficient. In Section V, we locate our
summary of results and final remarks. Finally, we place technical details in
Appendix A and Appendix B.

\section{Hamiltonian evolutions}

In this section, we present optimal and sub-optimal quantum Hamiltonian
evolutions linking arbitrary source and target states on the Bloch sphere.
Specifically, we study both evolutions by means of the notions of path length
\cite{wootters81,provost80,cafaro23}, geodesic efficiency
\cite{anandan90,cafaro20}, speed efficiency \cite{uzdin12}, and curvature
coefficients \cite{alsing24A,alsing24B} of the dynamical trajectories that
join the given initial and final states.

\subsection{Geodesic evolution}

The quantum evolution that specifies a geodesic evolution on the Bloch sphere
by evolving the initial state $\left\vert A\right\rangle \overset{\text{def}%
}{=}\left\vert \psi\left(  t_{A}\right)  \right\rangle $ into the final state
$\left\vert B\right\rangle \overset{\text{def}}{=}\left\vert \psi\left(
t_{B}\right)  \right\rangle $ in a time interval $t_{AB}\overset{\text{def}%
}{=}t_{B}-t_{A}$ is defined by the stationary Hamiltonian
\cite{brody06,brody07,bender07,bender09,ali09,cafaro23,rossetti24A,rossetti24B}%
\begin{equation}
\mathrm{H}_{\mathrm{opt}}\overset{\text{def}}{=}\mathbf{h}_{\mathrm{opt}}%
\cdot\mathbf{\boldsymbol{\sigma}}=E\frac{\hat{a}\times\hat{b}}{\sqrt{(\hat
{a}\times\hat{b})\cdot(\hat{a}\times\hat{b})}}\cdot\mathbf{\boldsymbol{\sigma
}}\text{,}\label{Hopt}%
\end{equation}
with $\mathbf{\boldsymbol{\sigma}}\overset{\text{def}}{\mathbf{\boldsymbol{=}%
}}\left(  \sigma_{x}\text{, }\sigma_{y}\text{, }\sigma_{z}\right)  $ being the
vector operator of Pauli matrices. We assume here to have $0\leq\theta
_{AB}\leq\pi$ with $\sqrt{(\hat{a}\times\hat{b})\cdot(\hat{a}\times\hat{b}%
)}=\sin\left(  \theta_{AB}\right)  $. The quantity $E$ denotes energy, while
the components of the vector $\mathbf{h}_{\mathrm{opt}}$ are measured in
energy units. However, we loosely denote $\mathbf{h}_{\mathrm{opt}}$ as the
\textquotedblleft magnetic\textquotedblright\ field vector. The unit vectors
$\hat{a}$ and $\hat{b}$ are the Bloch vectors that correspond to the states
$\left\vert A\right\rangle $ and $\left\vert B\right\rangle $, respectively,
with $\rho_{A}\overset{\text{def}}{=}(\mathbf{1+}\hat{a}\cdot
\mathbf{\boldsymbol{\sigma}})/2$ and $\rho_{B}\overset{\text{def}}%
{=}(\mathbf{1+}\hat{b}\cdot\mathbf{\boldsymbol{\sigma}})/2$. Clearly,
\textquotedblleft$\mathbf{1}$\textquotedblright\ represents the identity
operator here. The unitary evolution operator $U_{\mathrm{opt}}\left(
t\right)  $ that corresponds to $\mathrm{H}_{\mathrm{opt}}$ in Eq.
(\ref{Hopt}) is given by%
\begin{equation}
U_{\mathrm{opt}}\left(  t\right)  \overset{\text{def}}{=}\exp\left[  -\frac
{i}{\hslash}\mathrm{H}_{\mathrm{opt}}t\right]  =\cos\left(  \frac{E}{\hslash
}t\right)  \mathbf{1}-i\sin(\frac{E}{\hslash}t)\frac{\hat{a}\times\hat{b}%
}{\sin\left(  \theta_{AB}\right)  }\cdot\mathbf{\boldsymbol{\sigma}}%
\text{,}\label{Uopt}%
\end{equation}
with $\left\vert \psi(t)\right\rangle \overset{\text{def}}{=}U_{\mathrm{opt}%
}\left(  t\text{, }t_{A}\right)  \left\vert A\right\rangle =e^{-\frac
{i}{\hslash}\mathrm{H}_{\mathrm{opt}}(t-t_{A})}\left\vert A\right\rangle $ for
any $t_{A}\leq t\leq t_{B}$. The unitary time propagator defined by
$U_{\mathrm{opt}}\left(  t\right)  $ in Eq. (\ref{Uopt}) brings $\left\vert
A\right\rangle $ into $\left\vert B\right\rangle $ in a minimum time interval
given by $t_{AB}^{\mathrm{opt}}\overset{\text{def}}{=}\hslash\theta_{AB}%
/(2E)$, along a path of (shortest) length equal to $\theta_{AB}$. In terms of
the geodesic efficiency $\eta_{\mathrm{GE}}\overset{\text{def}}{=}s_{0}/s$ of
the quantum evolution \cite{anandan90,cafaro20},%
\begin{equation}
\eta_{\mathrm{GE}}\overset{\text{def}}{=}\frac{s_{0}}{s}=\frac{2\arccos\left[
\left\vert \left\langle A\left\vert B\right.  \right\rangle \right\vert
\right]  }{\int_{t_{A}}^{t_{B}}\frac{2}{\hslash}\Delta E(t)dt}\text{,}%
\label{NGE}%
\end{equation}
we have that $s_{0}=\theta_{AB}$, $t_{B}-t_{A}=t_{AB}^{\mathrm{opt}}%
\overset{\text{def}}{=}\hslash\theta_{AB}/(2E)$, $\Delta E^{2}\overset
{\text{def}}{=}\left\langle A\left\vert \mathrm{H}_{\mathrm{opt}}%
^{2}\right\vert A\right\rangle -\left\langle A\left\vert \mathrm{H}%
_{\mathrm{opt}}\right\vert A\right\rangle ^{2}=E^{2}$, and, therefore,
$s=\theta_{AB}$. Thus, $\eta_{\mathrm{GE}}^{\mathrm{opt}}=1$ and the evolution
is geodesic. Notice that $s_{0}$ denotes the distance along the shortest
geodesic path that joins $\left\vert A\right\rangle $ and $\left\vert
B\right\rangle $ on the complex projective Hilbert space. The quantity $s$,
instead, defines the distance along the actual dynamical trajectory
$\gamma\left(  t\right)  :t\mapsto\left\vert \psi\left(  t\right)
\right\rangle $ that corresponds to the evolution under $\mathrm{H}%
_{\mathrm{opt}}$ of the state vector $\left\vert \psi\left(  t\right)
\right\rangle $, with $t_{A}\leq t\leq t_{B}$. The evolution specified by
$U_{\mathrm{opt}}\left(  t\right)  $ in Eq. (\ref{Uopt}) occurs also without
any waste of energy resources since its speed efficiency \cite{uzdin12},%
\begin{equation}
\eta_{\mathrm{SE}}\overset{\text{def}}{=}\frac{\Delta\mathrm{H}_{\rho}%
}{\left\Vert \mathrm{H}\right\Vert _{\mathrm{SP}}}=\frac{\sqrt{\mathrm{tr}%
\left(  \rho\mathrm{H}^{2}\right)  -\left[  \mathrm{tr}\left(  \rho
\mathrm{H}\right)  \right]  ^{2}}}{\max\left[  \sqrt{\mathrm{eig}\left(
\mathrm{H}^{\dagger}\mathrm{H}\right)  }\right]  }\text{,}\label{NSE}%
\end{equation}
equals to one. This is a consequence of the fact that the energy uncertainty
$\Delta\mathrm{H}_{\rho}$, where $\rho=\rho\left(  t\right)  \overset
{\text{def}}{=}\left[  \mathbf{1+}\hat{a}\left(  t\right)  \cdot
\mathbf{\boldsymbol{\sigma}}\right]  /2$ and $t_{A}\leq t\leq t_{B}$ equals
the spectral norm $\left\Vert \mathrm{H}\right\Vert _{\mathrm{SP}}$ of the
Hamiltonian $\mathrm{H}_{\mathrm{opt}}$ in Eq. (\ref{Hopt}). Specifically,
from Eqs. (\ref{Hopt}) and (\ref{NSE}), we get%
\begin{equation}
\eta_{\mathrm{SE}}^{\mathrm{opt}}=\eta_{\mathrm{SE}}^{\mathrm{opt}}\left(
t\right)  \overset{\text{def}}{=}\frac{\sqrt{\mathbf{h}_{\mathrm{opt}}%
^{2}-(\hat{a}\cdot\mathbf{h}_{\mathrm{opt}})^{2}}}{\sqrt{\mathbf{h}%
_{\mathrm{opt}}^{2}}}=1\text{,}\label{se2}%
\end{equation}
since $\hat{a}\cdot\mathbf{h}_{\mathrm{opt}}=0$ for any instant $t\in\left[
t_{A}\text{, }t_{B}\right]  $. Therefore, a trajectory is wasteful in energy
(speed) whenever the magnetic field has a component along the Bloch vector,
thus generating useless rotations about $\hat{a}$. Finally, if we consider a
quantum evolution specified by a traceless Hamiltonian\textbf{ }%
$\mathrm{H}\overset{\text{def}}{=}\mathbf{h}\cdot\mathbf{\boldsymbol{\sigma}}%
$\textbf{, }with corresponding curvature coefficient given by
\cite{alsing24A,alsing24B}%
\begin{equation}
\kappa_{\mathrm{AC}}^{2}\left(  \hat{a}\text{, }\mathbf{h}\right)
\overset{\text{def}}{=}4\frac{\left(  \hat{a}\mathbf{\cdot h}\right)  ^{2}%
}{\mathbf{h}^{2}-\left(  \hat{a}\mathbf{\cdot h}\right)  ^{2}}\text{,}%
\label{way1}%
\end{equation}
we have%
\begin{equation}
\kappa_{\mathrm{AC}}^{2}\left(  \hat{a}\text{, }\mathbf{h}_{\mathrm{opt}%
}\right)  =0\text{,}\label{curva0}%
\end{equation}
since $\mathbf{h}_{\mathrm{opt}}$ is constantly orthogonal to $\hat{a}$. Given
the fact that $\kappa_{\mathrm{AC}}^{2}\left(  \hat{a}\text{, }\mathbf{h}%
\right)  $ in Eq. (\ref{way1}) is a measure of the bending of the \ path on
the Bloch sphere, a path with zero curvature exhibits no bending at all. For
more details on the concept of curvature for both stationary and nonstationary
Hamiltonian evolutions, we suggest Refs. \cite{alsing24A,alsing24B}. In
summary, the evolution specified by Eq. (\ref{Uopt}) occurs with unit geodesic
efficiency (i.e., with the shortest path length, that is, $\eta_{\mathrm{GE}%
}^{\mathrm{opt}}=1$), with unit speed efficiency (i.e., with no waste of
energy along the path, that is, $\eta_{\mathrm{SE}}^{\mathrm{opt}}\left(
t\right)  $ for any $t\in\left[  t_{A}\text{, }t_{B}\right]  $) and, finally,
with zero curvature coefficient (i.e., with no bending of the path, that is
$\left(  \kappa_{\mathrm{AC}}^{2}\right)  _{\mathrm{opt}}\overset{\text{def}%
}{=}\kappa_{\mathrm{AC}}^{2}\left(  \hat{a}\text{, }\mathbf{h}_{\mathrm{opt}%
}\right)  =0$).\begin{figure}[t]
\centering
\includegraphics[width=0.5\textwidth] {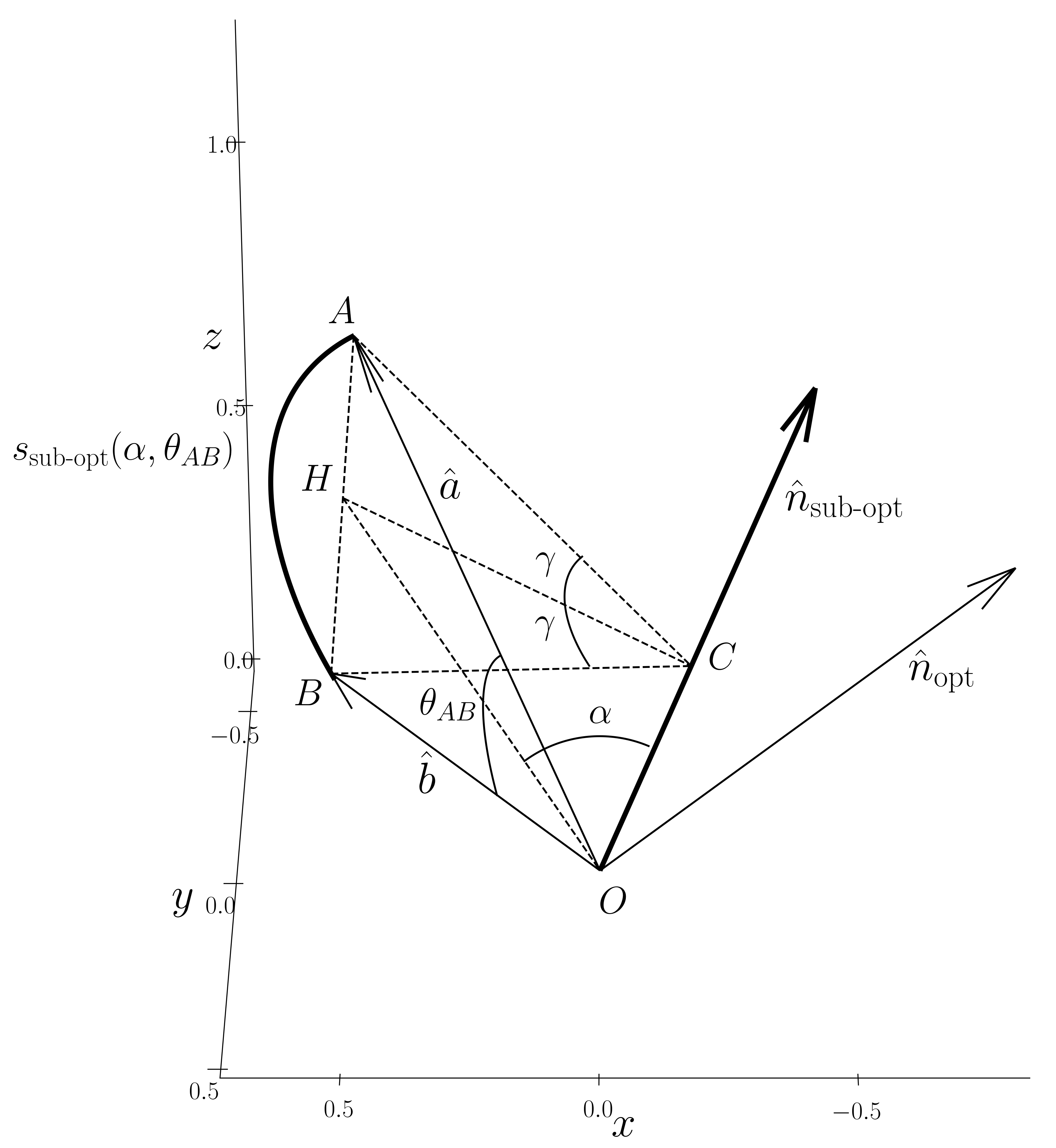}\caption{Illustration of the unit
vectors $\hat{n}_{\mathrm{opt}}\overset{\text{def}}{=}$ $(\hat{a}\times\hat
{b})/\left\Vert \hat{a}\times\hat{b}\right\Vert $ and $\hat{n}_{\mathrm{sub}%
\text{-}\mathrm{opt}}\overset{\text{def}}{=}\cos(\alpha)\left[  (\hat{a}%
+\hat{b})/\left\Vert \hat{a}+\hat{b}\right\Vert \right]  +\sin(\alpha)\left[
(\hat{a}\times\hat{b})/\left\Vert \hat{a}\times\hat{b}\right\Vert \right]  $
characterizing the optimal (thin black) and sub-optimal (thick black) rotation
axes, respectively. The initial and final unit Bloch vectors $\hat{a}$ and
$\hat{b}$, respectively, satisfy the condition $\hat{a}\cdot\hat{b}%
=\cos(\theta_{AB})$. The sub-optimal unitary evolution that defines the
transition from $\hat{a}$ to $\hat{b}$ is characterized by $\hat
{n}_{\mathrm{sub}\text{-}\mathrm{opt}}$ and the rotation angle $\phi
\overset{\text{def}}{=}2\gamma$. Finally, the length of the nongeodesic path
that connects these two Bloch vectors via the sub-optimal unitary evolution is
equal to $s_{\mathrm{sub}\text{-}\mathrm{opt}}\left(  \alpha\text{, }%
\theta_{AB}\right)  \overset{\text{def}}{=}\overline{AC}\left(  \alpha\text{,
}\theta_{AB}\right)  \phi\left(  \alpha\text{, }\theta_{AB}\right)  $ (thick
black).}%
\end{figure}

\subsection{Nongeodesic evolution}

The quantum evolution that describes a nongeodesic evolution on the Bloch
sphere by evolving the initial state $\left\vert A\right\rangle \overset
{\text{def}}{=}\left\vert \psi\left(  t_{A}\right)  \right\rangle $ into the
final state $\left\vert B\right\rangle \overset{\text{def}}{=}\left\vert
\psi\left(  t_{B}\right)  \right\rangle $ in a time interval $t_{AB}%
\overset{\text{def}}{=}t_{B}-t_{A}$ is specified by the stationary Hamiltonian
\cite{rossetti24B}%
\begin{equation}
\mathrm{H}_{\mathrm{sub}\text{\textrm{-}}\mathrm{opt}}\overset{\text{def}}%
{=}\mathbf{h}_{\mathrm{sub}\text{\textrm{-}}\mathrm{opt}}\cdot
\mathbf{\boldsymbol{\sigma}=}E\left[  \cos(\alpha)\frac{\hat{a}+\hat{b}}%
{\sqrt{(\hat{a}+\hat{b})\cdot(\hat{a}+\hat{b})}}+\sin(\alpha)\frac{\hat
{a}\times\hat{b}}{\sqrt{(\hat{a}\times\hat{b})\cdot(\hat{a}\times\hat{b})}%
}\right]  \cdot\mathbf{\boldsymbol{\sigma}}\text{,}\label{Hsopt}%
\end{equation}
with $0\leq\alpha\leq\pi$. Assuming to have $0\leq\theta_{AB}\leq\pi$,
$\sqrt{(\hat{a}\times\hat{b})\cdot(\hat{a}\times\hat{b})}=\sin\left(
\theta_{AB}\right)  $ and $\sqrt{(\hat{a}+\hat{b})\cdot(\hat{a}+\hat{b}%
)}=2\cos\left(  \theta_{AB}/2\right)  $. Therefore, the Hamiltonian in Eq.
(\ref{Hsopt}) can be recast as%
\begin{equation}
\mathrm{H}_{\mathrm{sub}\text{\textrm{-}}\mathrm{opt}}=E\left[  \cos
(\alpha)\frac{\hat{a}+\hat{b}}{2\cos\left(  \frac{\theta_{AB}}{2}\right)
}+\sin(\alpha)\frac{\hat{a}\times\hat{b}}{\sin\left(  \theta_{AB}\right)
}\right]  \cdot\mathbf{\boldsymbol{\sigma}}\text{.}%
\end{equation}
For a schematic depiction of the unit vectors $\hat{n}_{\mathrm{opt}}%
\overset{\text{def}}{=}\mathbf{h}_{\mathrm{sub}}/\sqrt{\mathbf{h}%
_{\mathrm{sub}}\cdot\mathbf{h}_{\mathrm{sub}}}$ and $\hat{n}_{\mathrm{sub}%
\text{-}\mathrm{opt}}\overset{\text{def}}{=}\mathbf{h}_{\mathrm{sub}%
\text{-}\mathrm{opt}}/\sqrt{\mathbf{h}_{\mathrm{sub}\text{-}\mathrm{opt}}%
\cdot\mathbf{h}_{\mathrm{sub}\text{-}\mathrm{opt}}}$ that specify the optimal
and sub-optimal Hamiltonians in Eqs. (\ref{Hopt}) and (\ref{Hsopt}),
respectively, we refer to Fig. $1$. In analogy to what we mentioned in the
optimal scenario case, the unit vectors $\hat{a}$ and $\hat{b}$ denote the
Bloch vectors that correspond to the states $\left\vert A\right\rangle $ and
$\left\vert B\right\rangle $, respectively, with $\rho_{A}\overset{\text{def}%
}{=}(\mathbf{1+}\hat{a}\cdot\mathbf{\boldsymbol{\sigma}})/2$ and $\rho
_{B}\overset{\text{def}}{=}(\mathbf{1+}\hat{b}\cdot\mathbf{\boldsymbol{\sigma
}})/2$. The unitary evolution operator $U_{\mathrm{sub}\text{-}\mathrm{opt}%
}\left(  t\right)  $ that corresponds to $\mathrm{H}_{\mathrm{sub}%
\text{\textrm{-}}\mathrm{opt}}$ in Eq. (\ref{Hsopt}) is given by%
\begin{equation}
U_{\mathrm{sub}\text{-}\mathrm{opt}}\left(  t\right)  \overset{\text{def}}%
{=}\exp\left[  -\frac{i}{\hslash}\mathrm{H}_{\mathrm{sub}\text{\textrm{-}%
}\mathrm{opt}}t\right]  =\cos\left(  \frac{E}{\hslash}t\right)  \mathbf{1}%
-i\sin(\frac{E}{\hslash}t)\left[  \cos(\alpha)\frac{\hat{a}+\hat{b}}%
{2\cos\left(  \frac{\theta_{AB}}{2}\right)  }+\sin(\alpha)\frac{\hat{a}%
\times\hat{b}}{\sin\left(  \theta_{AB}\right)  }\right]  \cdot
\mathbf{\boldsymbol{\sigma}}\text{,}\label{Usopt}%
\end{equation}
with $\left\vert \psi(t)\right\rangle \overset{\text{def}}{=}U_{\mathrm{sub}%
\text{-}\mathrm{opt}}\left(  t\text{, }t_{A}\right)  \left\vert A\right\rangle
=e^{-\frac{i}{\hslash}\mathrm{H}_{\mathrm{sub}\text{\textrm{-}}\mathrm{opt}%
}(t-t_{A})}\left\vert A\right\rangle $ for any $t_{A}\leq t\leq t_{B}$. The
unitary time propagator defined by $U_{\mathrm{sub}\text{-}\mathrm{opt}%
}\left(  t\right)  $ in Eq. (\ref{Uopt}) brings $\left\vert A\right\rangle $
into $\left\vert B\right\rangle $ in a sub-optimal time interval given by
$t_{AB}\left(  \alpha\right)  =t_{AB}^{\mathrm{sub}\text{-}\mathrm{opt}%
}\left(  \alpha\right)  $ along a path of length $s\left(  \alpha\right)  $.
The quantities $t_{AB}\left(  \alpha\right)  $ and $s\left(  \alpha\right)  $
are given by \cite{rossetti24B}
\begin{equation}
t_{AB}\left(  \alpha\right)  \overset{\text{def}}{=}\frac{\hbar}{E}%
\arccos{\left[  \sin({\alpha)}\frac{\cos({\frac{\theta_{AB}}{2})}}%
{\sqrt{1-\cos^{2}\left(  \alpha\right)  \cos^{2}\left(  \frac{\theta_{AB}}%
{2}\right)  }}\right]  }\text{,}\label{local}%
\end{equation}
and
\begin{equation}
s\left(  \alpha\right)  \overset{\text{def}}{=}2\sqrt{1-\cos^{2}\left(
\alpha\right)  \cos^{2}\left(  \frac{\theta_{AB}}{2}\right)  }\arccos{\left[
\sin({\alpha)}\frac{\cos({\frac{\theta_{AB}}{2})}}{\sqrt{1-\cos^{2}\left(
\alpha\right)  \cos^{2}\left(  \frac{\theta_{AB}}{2}\right)  }}\right]
}\text{,}\label{yoyo}%
\end{equation}
respectively. Since $s\left(  \alpha\right)  \geq s_{0}=\theta_{AB}$, the
geodesic efficiency $\eta_{\mathrm{GE}}$ in Eq. (\ref{NGE}) is less than one
in this sub-optimal case. Specifically, from Eqs. (\ref{NGE}) and
(\ref{yoyo}), we get%
\begin{equation}
\left(  \eta_{\mathrm{GE}}\right)  _{\mathrm{sub}\text{\textrm{-}}%
\mathrm{opt}}\left(  \alpha\right)  =\frac{\theta_{AB}}{2\sqrt{1-\cos
^{2}\left(  \alpha\right)  \cos^{2}\left(  \frac{\theta_{AB}}{2}\right)
}\arccos{\left[  \sin({\alpha)}\frac{\cos({\frac{\theta_{AB}}{2})}}%
{\sqrt{1-\cos^{2}\left(  \alpha\right)  \cos^{2}\left(  \frac{\theta_{AB}}%
{2}\right)  }}\right]  }}\text{.}\label{SOGE}%
\end{equation}
Furthermore, the evolution specified by $U_{\mathrm{sub}\text{-}\mathrm{opt}%
}\left(  t\right)  $ in Eq. (\ref{Uopt}) happens with some waste of energy
resources since its speed efficiency in Eq. (\ref{NSE}) is less than one in
this sub-optimal scenario. Indeed, decomposing $\mathbf{h}$ as $\mathbf{h}%
=\mathbf{h}_{\Vert}+\mathbf{h}_{\bot}=\left(  \mathbf{h\cdot}\hat{a}\right)
\hat{a}+\left[  \mathbf{h-}\left(  \mathbf{h\cdot}\hat{a}\right)  \hat
{a}\right]  $, for $\mathbf{h=h}_{\mathrm{sub}\text{\textrm{-}}\mathrm{opt}}$,
we get $\mathbf{h}_{\Vert}^{2}=E^{2}\cos^{2}\left(  \alpha\right)  \cos
^{2}(\theta_{AB}/2)$ and $\mathbf{h}_{\bot}^{2}=E^{2}\left[  1-\cos^{2}\left(
\alpha\right)  \cos^{2}(\theta_{AB}/2)\right]  $. Then, noting that the speed
efficiency of (traceless) stationary quantum evolutions can be recast as
$\eta_{\mathrm{SE}}=\sqrt{\mathbf{h}_{\bot}^{2}}/\sqrt{\mathbf{h}_{\bot}%
^{2}+\mathbf{h}_{\Vert}^{2}}$, we have%
\begin{equation}
\left(  \eta_{\mathrm{SE}}\right)  _{\mathrm{sub}\text{\textrm{-}}%
\mathrm{opt}}\left(  \alpha\right)  =\sqrt{1-\cos^{2}\left(  \alpha\right)
\cos^{2}\left(  \frac{\theta_{AB}}{2}\right)  }\text{.}\label{SOGE1}%
\end{equation}
Unlike the optimal case, in the sub-optimal case $\left(  \hat{a}%
\cdot\mathbf{h}_{\mathrm{sub}\text{\textrm{-}}\mathrm{opt}}\right)
^{2}=\mathbf{h}_{\Vert}^{2}\neq0$. For this reason there is some waste of
energy along the path and, as a consequence, the speed efficiency is not
maximal (i.e., not equal to one). Finally, given that the curvature
coefficient in Eq. (\ref{way1}) can be recast as $\kappa_{\mathrm{AC}}%
^{2}=4(\mathbf{h}_{\Vert}^{2}/\mathbf{h}_{\bot}^{2})$, we obtain%
\begin{equation}
\left(  \kappa_{\mathrm{AC}}^{2}\right)  _{\mathrm{sub}\text{\textrm{-}%
}\mathrm{opt}}\left(  \alpha\right)  =4\frac{\cos^{2}\left(  \alpha\right)
\cos^{2}\left(  \frac{\theta_{AB}}{2}\right)  }{1-\cos^{2}\left(
\alpha\right)  \cos^{2}\left(  \frac{\theta_{AB}}{2}\right)  }\text{.}%
\label{SOGE2}%
\end{equation}
In Fig. $2$, we plot the geodesic efficiency, the speed efficiency, and the
curvature coefficient in Eqs. (\ref{SOGE}), (\ref{SOGE1}), and (\ref{SOGE2}),
respectively, versus $\alpha\in\left[  0\text{, }\pi\right]  $ for
$\theta_{AB}=\pi/2$. Summing up, the quantum evolution defined in Eq.
(\ref{Usopt}) happens with sub-unit geodesic efficiency (i.e., without the
shortest path length, that is, $\left(  \eta_{\mathrm{GE}}\right)
_{\mathrm{sub}\text{\textrm{-}}\mathrm{opt}}$ in Eq. (\ref{SOGE}) is strictly
less than one), with sub-unit speed efficiency (i.e., with some waste of
energy along the path, that is, $\left(  \eta_{\mathrm{SE}}\right)
_{\mathrm{sub}\text{\textrm{-}}\mathrm{opt}}(t)$ in Eq. (\ref{SOGE1}) is
strictly less than one) and, finally, with nonzero curvature coefficient
(i.e., with some bending of the path, that is $\left(  \kappa_{\mathrm{AC}%
}^{2}\right)  _{\mathrm{sub}\text{\textrm{-}}\mathrm{opt}}\overset{\text{def}%
}{=}\kappa_{\mathrm{AC}}^{2}\left(  \hat{a}\text{, }\mathbf{h}_{\mathrm{sub}%
\text{\textrm{-}}\mathrm{opt}}\right)  $ in Eq. (\ref{SOGE2}) is strictly
greater than zero.

Having presented several aspects of the Hamiltonian evolutions being
considered in this paper, we are ready to introduce our proposed notions of
complexity and complexity length scale in the next section.\begin{figure}[t]
\centering
\includegraphics[width=0.75\textwidth] {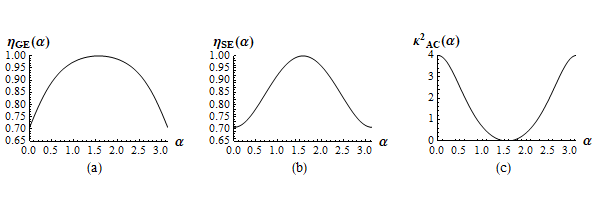}\caption{Plots of the geodesic
efficiency $\eta_{\mathrm{GE}}\left(  \alpha\right)  $, the speed efficiency
$\eta_{\mathrm{SE}}\left(  \alpha\right)  $, and the curvature coefficient
$\kappa_{\mathrm{AC}}^{2}\left(  \alpha\right)  $ versus $\alpha\in\left[
0\text{, }\pi\right]  $ in (a), (b), and (c), respectively. Note that for
$\alpha=\pi/2\simeq1.57$, $\eta_{\mathrm{GE}}=1$, $\eta_{\mathrm{SE}}=1$, and
$\kappa_{\mathrm{AC}}^{2}=0$.}%
\end{figure}

\section{Complexity}

In this section, to better motivate the introduction of our proposed concepts
of\emph{ complexity} and \emph{complexity length scale} of a quantum evolution
for a qubit state evolving under the action of an Hamiltonian that specifies a
time-dependent Schr\"{o}dinger equation, we briefly present our original
measure of complexity of motion on curved manifolds of probability
distributions in a probabilistic classical setting when only partial knowledge
about the system under investigation is available.

\subsection{Probabilistic classical setting}

We begin by recalling the concepts of information geometric complexity (IGC)
and information geometric entropy (IGE) \cite{cafaro07,cafarothesis}. Before
presenting formal definitions and more technical details, let us point out at
the beginning that the IGC is basically the exponential of the IGE. The
latter, in turn, represents the logarithm of the volume of the parametric
region accessed by the system during its (maximally probable, from an entropic
standpoint) geodesic evolution from an initial to a final point (i.e., a
probability distribution) on the underlying curved statistical manifold. The
IGE is a measure of complexity that was originally presented in Ref.
\cite{cafaro07} in the framework of the so-called Information Geometric
Approach to Chaos (IGAC) \cite{cafarothesis}. For a compact presentation of
the IGAC, we suggest Refs. \cite{ali18,ali21}.

In its original classical setting characterized by probability density
functions, the IGE can be introduced as follows. Suppose that the points
$\left\{  p\left(  x\text{; }\xi\right)  \right\}  $ of an $N_{\mathrm{c}}%
$-dimensional curved statistical manifold $\mathcal{M}_{s}$ are parametrized
by $N_{\mathrm{c}}$-real valued variables $\left(  \xi^{1}\text{,..., }%
\xi^{N_{\mathrm{c}}}\right)  $, with $\mathcal{M}_{s}$ defined as%
\begin{equation}
\mathcal{M}_{s}\overset{\text{def}}{=}\left\{  p\left(  x\text{; }\xi\right)
:\xi\overset{\text{def}}{=}\left(  \xi^{1}\text{,..., }\xi^{N_{\mathrm{c}}%
}\right)  \in\mathcal{D}_{\xi}^{\mathrm{tot}}\right\}  \text{.}\label{unooo}%
\end{equation}
Moreover, suppose that while the microvariables $x$ that characterize the
probability distributions $\left\{  p\left(  x\text{; }\xi\right)  \right\}  $
belong to the continuous microspace $\mathcal{X}$, the macrovariables $\xi$
are elements of the parameter space $\mathcal{D}_{\xi}^{\mathrm{tot}}$ given
by
\begin{equation}
\mathcal{D}_{\xi}^{\mathrm{tot}}\overset{\text{def}}{=}\left(  \mathcal{I}%
_{\xi^{1}}\otimes\mathcal{I}_{\xi^{2}}\text{...}\otimes\mathcal{I}%
_{\xi^{N_{\mathrm{c}}}}\right)  \subseteq\mathbb{R}^{N_{\mathrm{c}}}%
\text{.}\label{dtot}%
\end{equation}
Observe that $\mathcal{I}_{\xi^{k}}$ in Eq. (\ref{dtot}) denotes a subset of
$\mathbb{R}^{N}$ and specifies the range of acceptable values for the
statistical macrovariables $\xi^{k}$ where $1\leq k\leq N_{\mathrm{c}}$. The
IGE is a meaningful measure of temporal complexity of geodesic paths on
$\mathcal{M}_{s}$ in the IGAC context. It is formally defined as
\begin{equation}
\mathcal{S}_{\mathcal{M}_{s}}\left(  \tau\right)  \overset{\text{def}}{=}%
\log\widetilde{\mathrm{vol}}\left[  \mathcal{D}_{\xi}\left(  \tau\right)
\right]  \text{,}\label{IGE}%
\end{equation}
where the average dynamical statistical volume\textbf{\ }$\widetilde
{\mathrm{vol}}\left[  \mathcal{D}_{\xi}\left(  \tau\right)  \right]  $ is
given by
\begin{equation}
\widetilde{\mathrm{vol}}\left[  \mathcal{D}_{\xi}\left(  \tau\right)  \right]
\overset{\text{def}}{=}\frac{1}{\tau}\int_{0}^{\tau}\mathrm{vol}\left[
\mathcal{D}_{\xi}\left(  \tau^{\prime}\right)  \right]  d\tau^{\prime}%
\text{.}\label{rhs}%
\end{equation}
Since we take these volumes to be positive quantities, whenever we write
$\mathrm{vol}\left[  \mathcal{\cdot}\right]  $, we really mean $\left\vert
\mathrm{vol}\left[  \mathcal{\cdot}\right]  \right\vert \geq0$. Furthermore,
bear in mind that $\mathcal{D}_{\xi}\left(  \tau^{\prime}\right)  $ in Eq.
(\ref{rhs}) represents an $N_{\mathrm{c}}$-dimensional subspace of
$\mathcal{D}_{\xi}^{\mathrm{tot}}\subseteq\mathbb{R}^{N_{\mathrm{c}}}$ whose
elements $\left\{  \xi\right\}  $ with $\xi\overset{\text{def}}{=}\left(
\xi^{1}\text{,..., }\xi^{N_{\mathrm{c}}}\right)  $ fulfill $\xi^{k}\left(
\tau_{0}\right)  \leq\xi^{k}\leq\xi^{j}\left(  \tau_{0}+\tau^{\prime}\right)
$ where $\tau_{0}$ is \ the initial value assumed by the affine parameter
$\tau^{\prime}$ that specifies the geodesics on $\mathcal{M}_{s}$ as will be
explained in more depth soon. The tilde symbol in Eq. (\ref{rhs}) is employed
to denote the temporal average operation. Furthermore, we remark
that\textbf{\ }$\widetilde{\mathrm{vol}}\left[  \mathcal{D}_{\xi}\left(
\tau\right)  \right]  $\textbf{\ }in Eq. (\ref{rhs}) is determined by two
sequential integration procedures. The first integration is performed over the
accessed parameter space $\mathcal{D}_{\xi}\left(  \tau^{\prime}\right)  $ and
yields $\mathrm{vol}\left[  \mathcal{D}_{\xi}\left(  \tau^{\prime}\right)
\right]  $. The second integration, instead, yields $\widetilde{\mathrm{vol}%
}\left[  \mathcal{D}_{\xi}\left(  \tau\right)  \right]  $ and specifies a
temporal averaging procedure performed over the duration $\tau$ of the
geodesic evolution on the manifold $\mathcal{M}_{s}$. The quantity\textbf{\ }%
$\mathrm{vol}\left[  \mathcal{D}_{\xi}\left(  \tau^{\prime}\right)  \right]
$\textbf{\ }in the right-hand-side of Eq. (\ref{rhs}) denotes the volume of an
extended region on $\mathcal{M}_{s}$ and is defined as,%
\begin{equation}
\mathrm{vol}\left[  \mathcal{D}_{\xi}\left(  \tau^{\prime}\right)  \right]
\overset{\text{def}}{=}\int_{\mathcal{D}_{\xi}\left(  \tau^{\prime}\right)
}\rho\left(  \xi^{1}\text{,..., }\xi^{N}\right)  d^{N}\xi\text{,}\label{v}%
\end{equation}
with $\rho\left(  \xi^{1}\text{,..., }\xi^{N}\right)  $ being the so-called
Fisher density. It is equal to the square root of the determinant of the
Fisher-Rao information metric tensor $g_{\mu\nu}^{\mathrm{FR}}\left(
\xi\right)  $, that is, $\rho\left(  \xi\right)  \overset{\text{def}}{=}%
\sqrt{g^{\mathrm{FR}}\left(  \xi\right)  }$ with $g^{\mathrm{FR}}\left(
\xi\right)  \overset{\text{def}}{=}\det\left[  g_{\mu\nu}^{\mathrm{FR}}\left(
\xi\right)  \right]  $. Within the continuous microspace setting, recollect
that $g_{\mu\nu}^{\mathrm{FR}}\left(  \xi\right)  $ is given by%
\begin{equation}
g_{\mu\nu}^{\mathrm{FR}}\left(  \xi\right)  \overset{\text{def}}{=}\int
p\left(  x|\xi\right)  \partial_{\mu}\log p\left(  x|\xi\right)  \partial
_{\nu}\log p\left(  x|\xi\right)  dx\text{,}\label{FRmetric}%
\end{equation}
where $\partial_{\mu}\overset{\text{def}}{=}\partial/\partial\xi^{\mu}$ and
$1\leq\mu$, $\nu\leq N_{\mathrm{c}}$. Interestingly, observe that\textrm{
}$\mathrm{vol}\left[  \mathcal{D}_{\xi}\left(  \tau^{\prime}\right)  \right]
$ in Eq. (\ref{v}) acquires a simpler form when manifolds are provided with
metric tensors characterized by factorizable determinants such as,
\begin{equation}
g\left(  \xi\right)  =g\left(  \xi^{1}\text{,..., }\xi^{N_{\mathrm{c}}%
}\right)  \overset{\text{def}}{=}{\prod\limits_{k=1}^{N_{\mathrm{c}}}}%
g_{k}\left(  \xi^{k}\right)  \text{.}%
\end{equation}
In such a simplified (i.e., factorizable) scenario, the IGE in Eq. (\ref{IGE})
can be recast as
\begin{equation}
\mathcal{S}_{\mathcal{M}_{s}}\left(  \tau\right)  =\log\left\{  \frac{1}{\tau
}\int_{0}^{\tau}\left[  {\prod\limits_{k=1}^{N_{\mathrm{c}}}}\left(
\int_{\tau_{0}}^{\tau_{0}+\tau^{\prime}}\sqrt{g_{k}\left[  \xi^{k}\left(
\chi\right)  \right]  }\frac{d\xi^{k}}{d\chi}d\chi\right)  \right]
d\tau^{\prime}\right\}  \text{.}\label{IGEmod}%
\end{equation}
We stress that, when the microvariables $\left\{  x\right\}  $ are correlated,
$g\left(  \xi\right)  $ is not factorizable. In this more complicated (i.e.,
non-factorizable) scenario \cite{ali10}, one needs to employ the general
definition of the IGE. In the IGAC framework, the complexity of the
statistical models being studied can be specified by the leading asymptotic
behavior of $\mathcal{S}_{\mathcal{M}_{s}}\left(  \tau\right)  $ in Eq.
(\ref{IGEmod}). For this reason, we take into consideration the leading
asymptotic term in Eq. (\ref{IGE}) for the IGE,
\begin{equation}
\mathcal{S}_{\mathcal{M}_{s}}^{\text{\textrm{asymptotic}}}\left(  \tau\right)
\sim\lim_{\tau\rightarrow\infty}\left[  \mathcal{S}_{\mathcal{M}_{s}}\left(
\tau\right)  \right]  \text{.}\label{LONG}%
\end{equation}
The quantity $\mathcal{D}_{\xi}\left(  \tau^{\prime}\right)  $ defines the
domain of integration in the expression of $\mathrm{vol}\left[  \mathcal{D}%
_{\xi}\left(  \tau^{\prime}\right)  \right]  $ in Eq. (\ref{v}). It is given
by%
\begin{equation}
\mathcal{D}_{\xi}\left(  \tau^{\prime}\right)  \overset{\text{def}}{=}\left\{
\xi:\xi^{k}\left(  \tau_{0}\right)  \leq\xi^{k}\leq\xi^{k}\left(  \tau
_{0}+\tau^{\prime}\right)  \right\}  \text{,}\label{L1}%
\end{equation}
with $\tau_{0}\leq\chi\leq\tau_{0}+\tau^{\prime}$ and $\tau_{0}$ being the
initial value of the affine parameter $\chi$. In Eq. (\ref{L1}), $\xi^{k}%
=\xi^{k}\left(  \chi\right)  $ fulfills the geodesic equation
\begin{equation}
\frac{d^{2}\xi^{k}}{d\chi^{2}}+\Gamma_{ij}^{k}\frac{d\xi^{i}}{d\chi}\frac
{d\xi^{j}}{d\chi}=0\text{,}\label{ge}%
\end{equation}
where $\Gamma_{ij}^{k}$ in\ Eq. (\ref{ge}) are the Christoffel connection
coefficients,%
\begin{equation}
\Gamma_{ij}^{k}\overset{\text{def}}{=}\frac{1}{2}g^{kl}\left(  \partial
_{i}g_{lj}+\partial_{j}g_{il}-\partial_{l}g_{ij}\right)  \text{.}%
\end{equation}
Observe that $\mathcal{D}_{\xi}\left(  \tau^{\prime}\right)  $ in Eq.
(\ref{L1}) is an $N_{\mathrm{c}}$-dimensional subspace of $\mathcal{D}_{\xi
}^{\mathrm{tot}}$, with elements being $N_{\mathrm{c}}$-dimensional
macrovariables $\left\{  \xi\right\}  $ whose components $\xi^{j}$ are bounded
by fixed integration limits given by $\xi^{j}\left(  \tau_{0}\right)  $ and
$\xi^{j}\left(  \tau_{0}+\tau^{\prime}\right)  $. These limits can be found
via the integration of the $N_{\mathrm{c}}$-coupled nonlinear second order
ODEs in Eq. (\ref{ge}). Given the IGE, the so-called information geometric
complexity (IGC) $\mathcal{C}_{\mathcal{M}_{s}}\left(  \tau\right)  $ is
defined as
\begin{equation}
\mathcal{C}_{\mathcal{M}_{s}}\left(  \tau\right)  \overset{\text{def}}%
{=}\widetilde{\mathrm{vol}}\left[  \mathcal{D}_{\xi}\left(  \tau\right)
\right]  =e^{\mathcal{S}_{\mathcal{M}_{s}}\left(  \tau\right)  }%
\text{.}\label{IGC}%
\end{equation}
As previously stated, we shall set our attention on the asymptotic temporal
behavior of the IGC as given by $\mathcal{C}_{\mathcal{M}_{s}}%
^{\text{\textrm{asymptotic}}}\left(  \tau\right)  \overset{\tau\rightarrow
\infty}{\sim}e^{\mathcal{S}_{\mathcal{M}_{s}}\left(  \tau\right)  }$. The
interpretation of the IGE $\mathcal{S}_{\mathcal{M}_{s}}\left(  \tau\right)  $
in Eq. (\ref{IGE}) can be used to explain the significance of the IGC
$\mathcal{C}_{\mathcal{M}_{s}}\left(  \tau\right)  $. The IGE is a measure of
the number of the accessed macrostates in $\mathcal{M}_{s}$ defined in terms
of the temporal average of the\textbf{\ }$N_{\mathrm{c}}$\textbf{-}fold
integral of the Fisher density evaluated over geodesics viewed as maximum
probability trajectories. Specifically, the instantaneous IGE represents the
logarithm of the volume of the parameter space accessed by the system at that
particular instant in time. The temporal averaging technique in Eq.
(\ref{rhs}) is proposed to average out the possibly very intricate fine
details of the probabilistic dynamical description of the system on the
manifold $\mathcal{M}_{s}$. Moreover, the long-time limit in\ Eq. (\ref{LONG})
is employed to conveniently describe the identified dynamical indicators of
complexity by disregarding the transient effects which characterize the
calculation of the expected value of the volume of the accessed parameter space.

Summing up, the IGE offers an asymptotic coarse-grained inferential
description of the complex dynamics of a system in the presence of limited
available relevant information. For more in-depth information on the IGC and
the IGE, we suggest Refs. \cite{cafaro07,cafarothesis,cafaro17,cafaro18}.
Although we limited our discussion to a classical probabilistic continuous
setting, our approach can be easily extended to a classical probabilistic
discrete setting specified by probability mass functions. Additionally, with
some care, it can be extended to quantum probabilistic settings defined by
manifolds of parametrized density operators equipped with suitable metrics.

We are now ready to introduce our notions of \emph{complexity} and
\emph{complexity length scale} of a quantum evolution for a qubit state.

\subsection{Deterministic quantum setting}

In transitioning from the probabilistic classical setting to the deterministic
quantum setting for two-levels quantum systems, several differences occur.
First, we transition from an (entropic) geodesic motion on an arbitrary
$N_{\mathrm{c}}$-dimensional curved statistical manifold of probability
distributions to a (deterministic) quantum temporal evolution (either geodesic
or nongeodesic) on a two-dimensional Bloch sphere. Any dynamically evolving
unit pure quantum state on this sphere can be parametrized by means of two
real parameters, for instance the polar and the azimuthal angles. Second,
while the Fisher-Rao information metric is the natural metric on classical
statistical manifolds, the Fubini-Study metric is the natural metric on the
Bloch sphere. Third, the globally accessible parameter space in the quantum
domain is finite, unlike what generally happens in the classical probabilistic
domain. For instance, for a statistical manifold of Gaussian probability
distributions, the globally accessible parameter space in not finite. Fourth,
we expect fine details to be significantly less intricate than the ones that
occur in a classical probabilistic evolution. Nevertheless, the temporal
averaging technique can be equally important in the deterministic quantum
setting considered here in order to average out possible finer details and
keep track of a more relevant overall global behavior of the quantum
evolution. Fifth, the entropic nature of the evolution requires the study of
the asymptotic temporal behavior of statistical volumes to disregard temporary
transient effects and consider only long-lasting effects of the classical
probabilistic evolution. Due to the deterministic nature of the quantum
evolution that we take into consideration here, this long-time limit is no
longer required since there is no necessity to differentiate between
(nonpermanent) short-term and (permanent) long-term behaviors. Moreover, we
are considering here finite-time quantum processes for which the concept of
asymptotic temporal behavior is simply ill-defined.

\subsubsection{Complexity}

Given these preliminary remarks, to quantify the complex behavior of quantum
evolutions during a finite time-interval $\left[  t_{A}\text{, }t_{B}\right]
$, we propose the quantity \textrm{C}$\left(  t_{A}\text{, }t_{B}\right)  $
defined as%
\begin{equation}
\mathrm{C}\left(  t_{A}\text{, }t_{B}\right)  \overset{\text{def}}{=}%
\frac{\mathrm{V}_{\max}\left(  t_{A}\text{, }t_{B}\right)  -\overline
{\mathrm{V}}\left(  t_{A}\text{, }t_{B}\right)  }{\mathrm{V}_{\max}\left(
t_{A}\text{, }t_{B}\right)  }\text{.} \label{QCD}%
\end{equation}
In what follows, we specify the definitions of $\overline{\mathrm{V}}\left(
t_{A}\text{, }t_{B}\right)  $ and $\mathrm{V}_{\max}\left(  t_{A}\text{,
}t_{B}\right)  $ in Eq. (\ref{QCD}). Furthermore, we justify the reason why we
propose $\mathrm{C}\left(  t_{A}\text{, }t_{B}\right)  $ in Eq. (\ref{QCD}) as
a suitable measure of complexity.

To define the so-called \emph{accessed volume} $\overline{\mathrm{V}}\left(
t_{A}\text{, }t_{B}\right)  $, we schematically proceed as follows. Ideally,
integrate the time-dependent Schr\"{o}dinger evolution equation $i\hslash
\partial_{t}\left\vert \psi(t)\right\rangle =\mathrm{H}\left(  t\right)
\left\vert \psi(t)\right\rangle $ and express the (normalized) single-qubit
state vector $\left\vert \psi(t)\right\rangle $ at an arbitrary time $t$ in
terms of the computational basis state vectors $\left\{  \left\vert
0\right\rangle \text{, }\left\vert 1\right\rangle \right\}  $. We have
$\left\vert \psi(t)\right\rangle =c_{0}(t)\left\vert 0\right\rangle
+c_{1}(t)\left\vert 1\right\rangle $ with $c_{0}(t)$ and $c_{1}(t)$ given by
\begin{equation}
c_{0}(t)\overset{\text{def}}{=}\left\langle 0\left\vert \psi(t)\right.
\right\rangle =\left\vert c_{0}(t)\right\vert e^{i\phi_{0}(t)}\text{, and
}c_{1}(t)\overset{\text{def}}{=}\left\langle 1\left\vert \psi(t)\right.
\right\rangle =\left\vert c_{1}(t)\right\vert e^{i\phi_{1}(t)}\text{,}
\label{qa}%
\end{equation}
respectively. Furthermore, note that $\phi_{0}(t)$ and $\phi_{1}(t)$ are the
(real) phases of $c_{0}(t)$ and $c_{1}(t)$, respectively. Then, using the
complex quantum amplitudes $c_{0}(t)$ and $c_{1}(t)$ in Eq. (\ref{qa}), recast
the state $\left\vert \psi(t)\right\rangle $ in terms of a physically
equivalent state expressed in its canonical Bloch sphere representation
specified by the polar angle $\theta\left(  t\right)  \in\left[  0\text{, }%
\pi\right]  $ and the azimuthal angle $\varphi\left(  t\right)  \in\left[
0\text{, }2\pi\right)  $. From the temporal behavior of the two spherical
angles, construct the volume of the parametric region accessed by the quantum
system during its transition from $\left\vert \psi(t_{A})\right\rangle
=\left\vert A\right\rangle $ to $\left\vert \psi(t)\right\rangle $. Finally,
find the time-average volume of the parametric region accessed by the quantum
system during its transition from $\left\vert \psi(t_{A})\right\rangle
=\left\vert A\right\rangle $ to $\left\vert \psi(t_{B})\right\rangle
=\left\vert B\right\rangle $ with $t_{A}\leq t\leq t_{B}$.

After this preliminary outline, we go in some details here on how to calculate
$\overline{\mathrm{V}}\left(  t_{A}\text{, }t_{B}\right)  $. Using Eq.
(\ref{qa}), we observe that $\left\vert \psi(t)\right\rangle =c_{0}%
(t)\left\vert 0\right\rangle +c_{1}(t)\left\vert 1\right\rangle $ is
physically equivalent to the state $\left\vert c_{0}(t)\right\vert \left\vert
0\right\rangle +\left\vert c_{1}(t)\right\vert e^{i\left[  \phi_{1}%
(t)-\phi_{0}(t)\right]  }\left\vert 1\right\rangle $. Thus, $\left\vert
\psi(t)\right\rangle $ can be recast as
\begin{equation}
\left\vert \psi(t)\right\rangle =\cos\left(  \frac{\theta\left(  t\right)
}{2}\right)  \left\vert 0\right\rangle +e^{i\varphi\left(  t\right)  }%
\sin\left(  \frac{\theta\left(  t\right)  }{2}\right)  \left\vert
1\right\rangle \text{.}\label{qa2}%
\end{equation}
Formally, the polar angle $\theta\left(  t\right)  $ and the azimuthal angle
$\varphi\left(  t\right)  \overset{\text{def}}{=}\phi_{1}(t)-\phi_{0}%
(t)=\arg\left[  c_{1}(t)\right]  -\arg\left[  c_{0}(t)\right]  $ in Eq.
(\ref{qa2}) are given by,%
\begin{equation}
\theta\left(  t\right)  \overset{\text{def}}{=}2\arctan\left(  \frac
{\left\vert c_{1}(t)\right\vert }{\left\vert c_{0}(t)\right\vert }\right)
\text{,}\label{teta}%
\end{equation}
and, assuming $\operatorname{Re}\left[  c_{1}(t)\right]  >0$ and
$\operatorname{Re}\left[  c_{0}(t)\right]  >0$,
\begin{equation}
\varphi\left(  t\right)  \overset{\text{def}}{=}\arctan\left\{  \frac
{\operatorname{Im}\left[  c_{1}(t)\right]  }{\operatorname{Re}\left[
c_{1}(t)\right]  }\right\}  -\arctan\left\{  \frac{\operatorname{Im}\left[
c_{0}(t)\right]  }{\operatorname{Re}\left[  c_{0}(t)\right]  }\right\}
\text{,}\label{fi}%
\end{equation}
respectively. More generally, the expression for $\varphi\left(  t\right)  $
in Eq. (\ref{fi}) can become more complicated. This is due to the fact that,
in general, the phase $\arg\left(  z\right)  $ of a complex number $%
%TCIMACRO{\U{2102} }%
%BeginExpansion
\mathbb{C}
%EndExpansion
\ni z\overset{\text{def}}{=}x+iy=\left\vert z\right\vert e^{i\arg(z)}$ must be
expressed by means of the $2$-\textrm{argument arctangent} function
\textrm{atan}$2$ as $\arg(z)=$\textrm{atan}$2(y$, $x)$. When $x>0$,
\textrm{atan}$2(y$, $x)$ reduces to $\arctan\left(  y/x\right)  $. For more
details, we refer to Ref. \cite{grad00}. Then, the unit Bloch vector $\hat
{r}\left(  t\right)  $ that corresponds to the state vector $\left\vert
\psi(t)\right\rangle $ in Eq. (\ref{qa2}) equals $\hat{r}\left(  t\right)
=(\sin\left[  \theta\left(  t\right)  \right]  \cos\left[  \varphi\left(
t\right)  \right]  $, $\sin\left[  \theta\left(  t\right)  \right]
\sin\left[  \varphi\left(  t\right)  \right]  $, $\cos\left[  \theta\left(
t\right)  \right]  )$. We are now in a position where we can define
$\overline{\mathrm{V}}\left(  t_{A}\text{, }t_{B}\right)  $. Namely, the
accessed volume $\overline{\mathrm{V}}\left(  t_{A}\text{, }t_{B}\right)  $
that corresponds to the quantum evolution from $\left\vert \psi(t_{A}%
)\right\rangle =\left\vert A\right\rangle $ to $\left\vert \psi(t_{B}%
)\right\rangle =\left\vert B\right\rangle $, with $t_{A}\leq t\leq t_{B}$,
under the Hamiltonian \textrm{H}$\left(  t\right)  $\textbf{, }is defined as%
\begin{equation}
\overline{\mathrm{V}}\left(  t_{A}\text{, }t_{B}\right)  \overset{\text{def}%
}{=}\frac{1}{t_{B}-t_{A}}\int_{t_{A}}^{t_{B}}V(t)dt\text{.}%
\label{avgcomplexity}%
\end{equation}
The quantity $V(t)$ in Eq. (\ref{avgcomplexity}) is the instantaneous volume
given by,%
\begin{equation}
V(t)=V(\theta(t)\text{, }\varphi\left(  t\right)  )\overset{\text{def}}%
{=}\mathrm{vol}\left[  \mathcal{D}_{\mathrm{accessed}}\left[  \theta(t)\text{,
}\varphi(t)\right]  \right]  \text{,}\label{local-complexity}%
\end{equation}
with $\mathrm{vol}\left[  \mathcal{D}_{\mathrm{accessed}}\left[
\theta(t)\text{, }\varphi(t)\right]  \right]  $ being defined as,%
\begin{equation}
\mathrm{vol}\left[  \mathcal{D}_{\mathrm{accessed}}\left[  \theta(t)\text{,
}\varphi(t)\right]  \right]  \overset{\text{def}}{=}\int\int_{\mathcal{D}%
_{\mathrm{accessed}}\left[  \theta(t)\text{, }\varphi(t)\right]  }%
\sqrt{g_{\mathrm{FS}}\left(  \theta\text{, }\varphi\right)  }d\theta
d\varphi\text{.}\label{q3}%
\end{equation}
We reiterate that we take these volumes to be positive quantities. Therefore,
whenever we write $\mathrm{vol}\left[  \mathcal{\cdot}\right]  $, we really
mean $\left\vert \mathrm{vol}\left[  \mathcal{\cdot}\right]  \right\vert
\geq0$. In Eq. (\ref{q3}), $g_{\mathrm{FS}}\left(  \theta\text{, }%
\varphi\right)  \overset{\text{def}}{=}\sqrt{\sin^{2}(\theta)/16}$ is the
determinant of the matrix that corresponds to the Fubini-Study infinitesimal
line element $ds_{\mathrm{FS}}^{2}\overset{\text{def}}{=}(1/4)\left[
d\theta^{2}+\sin^{2}(\theta)d\varphi^{2}\right]  $. Finally, $\mathcal{D}%
_{\mathrm{accessed}}\left[  \theta(t)\text{, }\varphi(t)\right]  $ in Eq.
(\ref{q3}) specifies the parametric region accessed by the quantum system
during its transition from the initial state $\left\vert \psi(t_{A}%
)\right\rangle =\left\vert A\right\rangle $ to an intermediate state
$\left\vert \psi(t)\right\rangle $, with $t_{A}\leq t\leq t_{B}$. It is
defined as%
\begin{equation}
\mathcal{D}_{\mathrm{accessed}}\left[  \theta(t)\text{, }\varphi(t)\right]
\overset{\text{def}}{=}\left[  \theta\left(  t_{A}\right)  \text{, }%
\theta\left(  t\right)  \right]  \times\left[  \varphi\left(  t_{A}\right)
\text{, }\varphi\left(  t\right)  \right]  \subset\left[  0\text{, }%
\pi\right]  _{\theta}\times\left[  0\text{, }2\pi\right)  _{\varphi}%
\text{.}\label{j5B}%
\end{equation}
For computational convenience, we remark that the instantaneous volume $V(t)$
in Eq. (\ref{local-complexity}) can be simply expressed as $V(t)=\left\vert
\left(  \cos\left[  \theta\left(  t_{A}\right)  \right]  -\cos\left[
\theta\left(  t\right)  \right]  \right)  \left(  \varphi(t)-\varphi
(t_{A})\right)  \right\vert /4$ with $\theta\left(  t\right)  $ and
$\varphi\left(  t\right)  $ in Eqs. (\ref{teta}) and (\ref{fi}), respectively.
In conclusion, inspired by our classical probabilistic setting (where the
tilde symbol is used to denote the time-average procedure), the accessed
volume $\overline{\mathrm{V}}\left(  t_{A}\text{, }t_{B}\right)  $ in Eq.
(\ref{avgcomplexity}) can be recast as%
\begin{equation}
\overline{\mathrm{V}}\left(  t_{A}\text{, }t_{B}\right)  \overset{\text{def}%
}{=}\widetilde{\mathrm{vol}}\left[  \mathcal{D}_{\mathrm{accessed}}\left[
\theta(t)\text{, }\varphi(t)\right]  \right]  \text{,}\label{cafe1}%
\end{equation}
with $t_{A}\leq t\leq t_{B}$ in Eq. (\ref{cafe1}).

To introduce the so-called \emph{accessible volume} $\mathrm{V}_{\max}\left(
t_{A}\text{, }t_{B}\right)  $, we need to present some preliminary remarks.
The departure, with respect to the classical probabilistic setting, from the
focus on the asymptotic temporal behavior of the complexity measure together
with the boundedness of the globally accessible parametric regions demand some
care. Indeed, these two aspects, along with the fact that we are now
interested in comparing the complexity of distinct quantum evolutions
connecting the same pair of initial and final states, lead us to consider a
notion of quantum complexity length scale.

We recall that the idea that complexity is more than length appears in
different fields of science, including computer science
\cite{beizer84,gill91,ande94}\ and evolutionary biology \cite{bonner04}. For
instance, when defining the concept of software complexity in computer
science, one can argue that the complexity should be an attribute that
measures something independent of the length of a software program
\cite{ande94}. For example, complexity might increase with the size of the
task, but this increase should usually be less than the increase in size.
Furthermore, when defining the notion of evolutionary complexity in biology,
one observes that there can be large variations in the so-called
size-complexity rule \cite{bonner04}. Indeed, the relationship between size
(or, alternatively, length) and complexity can be quite complicated when
investigating the evolution of biological systems such as small multicellular
organisms. For such systems, for example, an increase or decrease in
complexity does not necessarily require changes in size. Finally, for an
interesting discussion on how complexity should scale with size in an
information theory characterization of physical systems, we suggest Ref.
\cite{ay08}. Then, keeping in mind that complexity is more than length, we
propose to construct a quantum complexity length scale as a product of two
main factors. The first factor is the actual length of the path on the Bloch
sphere that connects the given initial and final states. Clearly, different
Hamiltonians yield different paths with different lengths during the evolution
in a finite time interval $t_{B}-t_{A}$. The second factor, instead, takes
into account the fact that different Hamiltonians give rise to different
bounded accessed volumes of parametric regions where each one can be
\textquotedblleft boxed\textquotedblright\ with a corresponding greater (yet
bounded) local accessible volume. The ratio between these two volumes, in
turn, can be regarded as a sort of normalized (local) volume ratio that will
specify the second factor in our proposed quantum complexity length scale
discussed in the next subsection. Given this set of remarks, we define the
accessible volume $\mathrm{V}_{\max}(t_{A}$, $t_{B})$ as%
\begin{equation}
\mathrm{V}_{\max}(t_{A}\text{, }t_{B})\overset{\text{def}}{=}\mathrm{vol}%
\left[  \mathcal{D}_{\text{\textrm{accessible}}}\left(  \theta\text{, }%
\varphi\right)  \right]  =\int\int_{\mathcal{D}_{\text{\textrm{accessible}}%
}\left(  \theta\text{, }\varphi\right)  }\sqrt{g_{\mathrm{FS}}\left(
\theta\text{, }\varphi\right)  }d\theta d\varphi\text{.}\label{j4}%
\end{equation}
The quantity $\mathcal{D}_{\text{\textrm{accessible}}}\left(  \theta\text{,
}\varphi\right)  $ in Eq. (\ref{j4}) is the (local) maximally accessible
two-dimensional parametric region during the quantum evolution from
$\left\vert \psi_{A}\left(  \theta_{A}\text{, }\varphi_{A}\right)
\right\rangle $ to $\left\vert \psi\left(  \theta_{B}\text{, }\varphi
_{B}\right)  \right\rangle $ and is defined as%
\begin{equation}
\mathcal{D}_{\text{\textrm{accessible}}}\left(  \theta\text{, }\varphi\right)
\overset{\text{def}}{=}\left\{  \left(  \theta\text{, }\varphi\right)
:\theta_{\min}\leq\theta\leq\theta_{\max}\text{, and }\varphi_{\min}%
\leq\varphi\leq\varphi_{\max}\right\}  \text{.}\label{j5}%
\end{equation}
Note that $\theta_{\min}$, $\theta_{\max}$, $\varphi_{\min}$, and
$\varphi_{\max}$ in Eq. (\ref{j5}) are given by
\begin{equation}
\theta_{\min}\overset{\text{def}}{=}\underset{t_{A}\leq t\leq t_{B}}{\min
}\theta(t)\text{, }\theta_{\max}\overset{\text{def}}{=}\underset{t_{A}\leq
t\leq t_{B}}{\max}\theta(t)\text{, }\varphi_{\min}\overset{\text{def}}%
{=}\underset{t_{A}\leq t\leq t_{B}}{\min}\varphi(t)\text{, and }\varphi_{\max
}\overset{\text{def}}{=}\underset{t_{A}\leq t\leq t_{B}}{\max}\varphi
(t)\text{,}\label{minmax}%
\end{equation}
respectively. Furthermore, observe that we have $\mathcal{D}%
_{\text{\textrm{accessed}}}\left(  \theta\text{, }\varphi\right)
\subset\mathcal{D}_{\text{\textrm{accessible}}}\left(  \theta\text{, }%
\varphi\right)  \subset\left[  0,\pi\right]  _{\theta}\times\left[  0\text{,
}2\pi\right)  _{\varphi}$. Finally, given $\overline{\mathrm{V}}\left(
t_{A}\text{, }t_{B}\right)  $ and $\mathrm{V}_{\max}\left(  t_{A}\text{,
}t_{B}\right)  $ in Eqs. (\ref{avgcomplexity}) and (\ref{j4}), respectively,
our proposed concept of complexity \textrm{C}$\left(  t_{A}\text{, }%
t_{B}\right)  $ in Eq. (\ref{QCD}) is formally defined. It is worthwhile
pointing out here that in the context of Bayesian model selection, the
so-called Bayesian complexity is formally defined as the natural logarithm of
the ratio between two volumes, \textrm{C}$_{\mathrm{Bayesian}}\overset
{\text{def}}{=}\ln\left[  V(f)/V_{\mathrm{c}}(f)\right]  $ \cite{myung00}. The
quantities $V_{\mathrm{c}}(f)$ and $V(f)$ denote the volume of the
distinguishable distributions in $f$ that come close to the truth and the
total volume of the model family $f$, respectively. In particular, a small
ratio $V_{\mathrm{c}}(f)/V(f)$ specifies (complex) models that occupy a small
volume close to the truth relative to the total volume of the model. For this
reason, a complex model (in a Bayesian sense) is one with a small fraction of
its distinguishable probability distributions lying near the truth
\cite{myung00}. Similarly, identifying $V_{\mathrm{c}}\leftrightarrow
\overline{\mathrm{V}}$ and $V\leftrightarrow\mathrm{V}_{\max}$, the ratio
$\overline{\mathrm{V}}/\mathrm{V}_{\max}$ is introduced to compare in a
meaningful way different Hamiltonian evolutions. These, in turn, can be
possibly specified by parametric regions of different dimension\textbf{.}
Furthermore, the ratio $\overline{\mathrm{V}}/\mathrm{V}_{\max}$ penalizes
Hamiltonian evolutions with a corresponding accessed (i.e., occupied)
parametric region of volume equal to only a small fraction of the accessible
volume. Lastly, a complex quantum Hamiltonian evolution is one with a small
fraction $\overline{\mathrm{V}}/\mathrm{V}_{\max}$, i.e. with a large fraction
$\left(  \mathrm{V}_{\max}-\overline{\mathrm{V}}\right)  /\mathrm{V}_{\max}$.

For clarity, it is worthwhile pointing out that the\textbf{ }\textrm{C}%
$\left(  t_{A}\text{, }t_{B}\right)  $ in Eq. (\ref{QCD}) is well-defined for
a quantum evolution that occurs along a meridian (specified by a constant
$\varphi$) or a parallel (characterized by a constant $\theta$). For clarity
of exposition, consider the\textbf{ }evolution from an initial state
$\left\vert A\right\rangle \overset{\text{def}}{=}\left[  \left\vert
0\right\rangle +\left\vert 1\right\rangle \right]  /\sqrt{2}$ to a final state
$\left\vert B\right\rangle \overset{\text{def}}{=}$ $\left\vert A\right\rangle
=\left[  \left\vert 0\right\rangle +i\left\vert 1\right\rangle \right]
/\sqrt{2}$ under an Hamiltonian of the form \textrm{H}$\overset{\text{def}}%
{=}E\sigma_{z}$ with $E\overset{\text{def}}{=}\hslash\omega$. During this
evolution from $t_{A}=0$ to $t_{B}=\pi/(4\omega)$, $\theta\left(  t\right)
=\pi/2$ (i.e., a parallel on the Bloch sphere) and $\varphi\left(  t\right)
=2\omega t$ for any $t_{A}\leq t\leq t_{B}$. Furthermore, a simple calculation
yields $\overline{\mathrm{V}}\left(  t_{A}\text{, }t_{B}\right)  =\pi/8$ and
\textrm{V}$_{\max}\left(  t_{A}\text{, }t_{B}\right)  =\pi/4$. Thus,\textbf{
}\textrm{C}$\left(  t_{A}\text{, }t_{B}\right)  =1/2$. A consistency check, we
emphasize that we expect to obtain the same complexity for a quantum evolution
from an initial state $\left\vert A\right\rangle \overset{\text{def}}%
{=}\left[  \left\vert 0\right\rangle +i\left\vert 1\right\rangle \right]
/\sqrt{2}$\textbf{ }to a final state $\left\vert B\right\rangle \overset
{\text{def}}{=}$ $\left\vert 0\right\rangle $ under an Hamiltonian of the form
\textrm{H}$\overset{\text{def}}{=}E\sigma_{x}$ with $E\overset{\text{def}}%
{=}\hslash\omega$. During this evolution from\textbf{ }$t_{A}=0$\textbf{ }to
$t_{B}=\pi/(4\omega)$, $\theta\left(  t\right)  =2\arctan\left[  \sqrt{\left(
\frac{\cos(\omega t)-\sin\left(  \omega t\right)  }{\cos(\omega t)+\sin\left(
\omega t\right)  }\right)  ^{2}}\right]  $ and $\varphi\left(  t\right)
=\pi/2$ (i.e., a meridian on the Bloch sphere) for any $t_{A}\leq t\leq t_{B}%
$. Again, a simple calculation leads to\textbf{ }$\overline{\mathrm{V}}\left(
t_{A}\text{, }t_{B}\right)  =\pi/8$ and \textrm{V}$_{\max}\left(  t_{A}\text{,
}t_{B}\right)  =\pi/4$. Thus, as expected on physical grounds, \textrm{C}%
$\left(  t_{A}\text{, }t_{B}\right)  =1/2$ as in the previous illustrative
example.\textbf{ }

In Fig.\textbf{ }$3$\textbf{, }we provide an intuitive picture to help
grasping the meaning of the concept of complexity proposed in Eq.
(\ref{QCD}).\begin{figure}[t]
\centering
\includegraphics[width=0.6\textwidth] {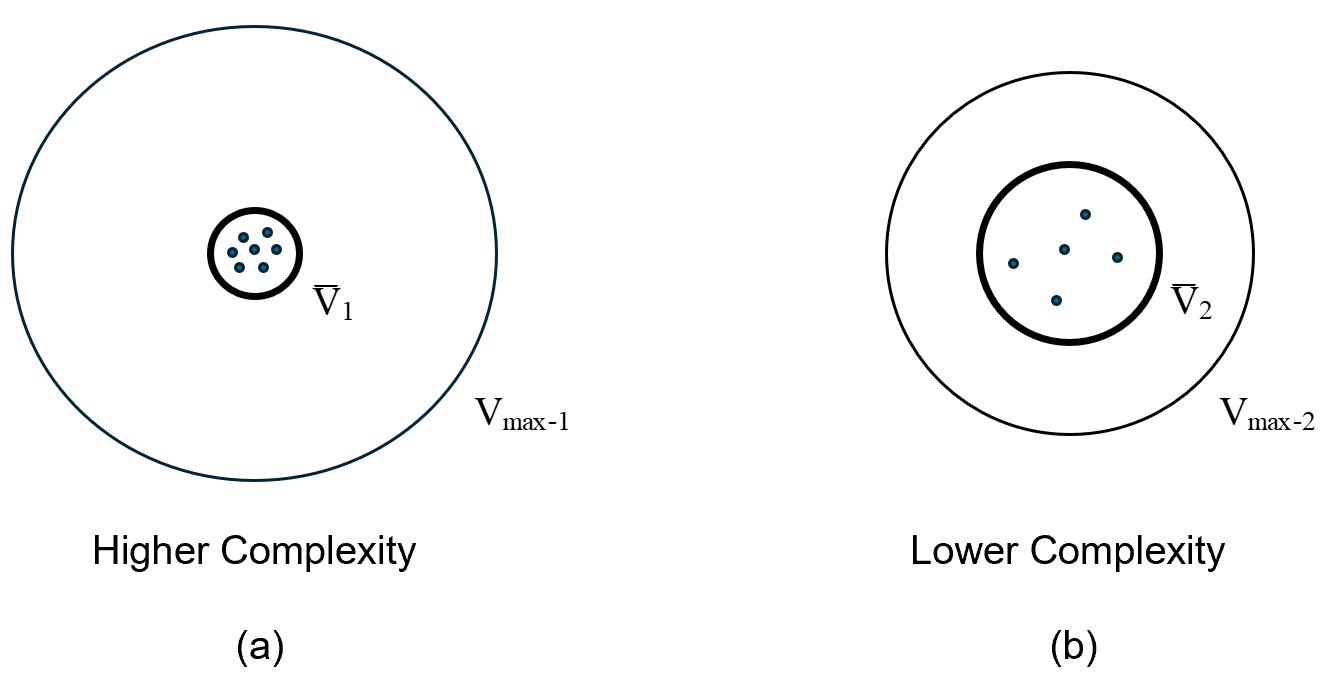}\caption{Schematic depiction that
helps grasping the concept of complexity \textrm{C} in Eq. (\ref{QCD}). In
(a), we sketch a complex scenario where the accessed volume $\overline
{\mathrm{V}}_{1}$\textrm{\ }(region enclosed by the thick circular line) is a
small fraction of the accessible volume \textrm{V}$_{\max\text{-}1}$ (region
enclosed by the thin circular line). In (b), instead, we describe a scenario
with a smaller degree of complexity since $\overline{\mathrm{V}}_{2}$ (region
enclosed by the thick circular line) is a larger fraction of \textrm{V}%
$_{\max\text{-}2}$ (region enclosed by the thin circular line). In general, to
a higher complexity scenario there corresponds a long quantum path of length
$s_{1}$ specified by a small fraction ($\overline{\mathrm{V}}_{1}%
/\mathrm{V}_{\max\text{-}1}$) of the accessible volume \textrm{V}%
$_{\max\text{-}1}$. To a lower complexity scenario, instead, there corresponds
a short quantum path of length $s_{2}<s_{1}$ described by a large fraction
($\overline{\mathrm{V}}_{2}/\mathrm{V}_{\max\text{-}2}$, with $\overline
{\mathrm{V}}_{2}/\mathrm{V}_{\max\text{-}2}>\overline{\mathrm{V}}%
_{1}/\mathrm{V}_{\max\text{-}1}$) of the accessible volume \textrm{V}%
$_{\max\text{-}2}$. The number of points depicted in an accessed volume
$\overline{\mathrm{V}}$ is taken here to be proportional to the squared length
of the path $s^{2}$ amplified by a (greater than one) factor which is the
reciprocal of $\overline{\mathrm{V}}/\mathrm{V}_{\max}$ with $0\leq$
$\overline{\mathrm{V}}/\mathrm{V}_{\max}\leq1$. These two scenarios are
well-described by the notion of complexity length scale \textrm{L}%
$_{\mathrm{C}}$ in\ Eq. (\ref{complexityL}), with \textrm{L}$_{\mathrm{C}}%
^{2}=s^{2}\cdot(\overline{\mathrm{V}}/\mathrm{V}_{\max})^{-1}$.}%
\end{figure}

Having proposed a notion of complexity for a quantum evolution, we can
introduce now the concept of complexity length scale.

\subsubsection{Complexity length scale}

We begin by recalling the concept of length of a path on the Bloch sphere. To
describe the physical significance of the Riemannian distance between two
arbitrarily selected pure quantum states, we suggest the work by Wootters in
Ref. \cite{wootters81}. For mixed states, instead, we suggest the analysis
carried out by Braunstein and Caves in Ref. \cite{braunstein94}. Given two
infinitesimally close points $\xi$ and $\xi+d\xi$ along a path $\xi\left(
\chi\right)  $ with $\chi_{1}\leq\chi\leq\chi_{2}$, they can be considered
statistically distinguishable if $d\xi$ is greater than (or equal to) the
standard fluctuation of $\xi$ \cite{diosi84}. In general, the infinitesimal
line element along the path equals $ds_{\mathrm{FS}}$ with $ds_{\mathrm{FS}%
}^{2}=g_{\mu\nu}^{\mathrm{FS}}\left(  \xi\right)  d\xi^{\mu}d\xi^{\nu}$ with
$1\leq\mu$, $\nu\leq m$, where $m$ represents the number of real parameters
used to parametrize the quantum state. The length $\mathcal{L}$ of the path
$\xi\left(  \chi\right)  $ with $\chi_{1}\leq\chi\leq\chi_{2}$ between
$\xi_{1}\overset{\text{def}}{=}\xi\left(  \chi_{1}\right)  $ and $\xi
_{2}\overset{\text{def}}{=}\xi\left(  \chi_{2}\right)  $ is given by%
\begin{equation}
\mathcal{L}\overset{\text{def}}{=}\int_{\xi_{1}}^{\xi_{2}}\sqrt
{ds_{\mathrm{FS}}^{2}}=\int_{\chi_{1}}^{\chi_{2}}\sqrt{\frac{ds_{\mathrm{FS}%
}^{2}}{d\chi^{2}}}d\chi\text{,}%
\end{equation}
and denotes the maximal number $\tilde{N}$ of statistically distinguishable
quantum states along the path $\xi\left(  \chi\right)  $. For two-level
quantum systems, $ds_{\mathrm{FS}}^{2}\overset{\text{def}}{=}(1/4)\left[
d\theta^{2}+\sin^{2}(\theta)d\varphi^{2}\right]  $. Interestingly, the
geodesic distance between $\xi_{1}$ and $\xi_{2}$ represents the path of
shortest distance between $\xi_{1}$ and $\xi_{2}$ and is given by the minimum
of $\tilde{N}$. This link between path length and number of statistically
distinguishable states along the path is a significant physical consideration
to keep in mind throughout our analysis. Besides, this perspective can be
generalized in a natural fashion to the geometric characterization of quantum
mixed states \cite{braunstein94}. More explicitly, from a geodesic efficiency
standpoint, we know that an efficient quantum evolution is the one with a
length of the path $s$ given by \cite{anandan90},%
\begin{equation}
s\overset{\text{def}}{=}\int_{t_{A}}^{t_{B}}2\frac{\Delta E\left(  t\right)
}{\hslash}dt\text{,}\label{j3}%
\end{equation}
connecting an initial (i.e., $\left\vert \psi_{A}\left(  \theta_{A}\text{,
}\varphi_{A}\right)  \right\rangle $) and a final (i.e., $\left\vert \psi
_{B}\left(  \theta_{B}\text{, }\varphi_{B}\right)  \right\rangle $) quantum
state as small as possible (that is, with the length as close as possible to
the shortest geodesic path length $s_{0}\overset{\text{def}}{=}2\arccos\left[
\left\vert \left\langle \psi_{A}\left(  \theta_{A}\text{, }\varphi_{A}\right)
\left\vert \psi_{B}\left(  \theta_{B}\text{, }\varphi_{B}\right)  \right.
\right\rangle \right\vert \right]  $ connecting the two states). The quantity
$\Delta E\left(  t\right)  $ in Eq. (\ref{j3}) is the energy uncertainty
(i.e., the square-root of the variance of the Hamiltonian operator with
respect to the state $\left\vert \psi\left(  t\right)  \right\rangle $).
Therefore, ranking quantum evolutions in terms of the geodesic efficiency
$\eta_{\mathrm{GE}}\overset{\text{def}}{=}s_{0}/s$ \cite{anandan90,cafaro20})
is straightforward. At the same time, one might ask if it is reasonable to
expect that the most efficient quantum evolution occurs by exploring the
(smallest) average volume of the accessed parametric region in the shortest
possible time. Unlike what one might be tempted to think from the classical
probabilistic approach, it turns out that ranking the performance of quantum
evolutions is more subtle than expected when one takes into account the
average volume of the accessed parametric region (i.e., the accessed volume
$\overline{\mathrm{V}}(t_{A}$, $t_{B})$ in Eq. (\ref{avgcomplexity})) together
with the maximum volume of the accessible parametric region specified by the
accessible volume \textrm{V}$_{\max}(t_{A}$, $t_{B})$ in Eq. (\ref{j4}).

In the spirit that complexity is more than length, we express in a
quantitative manner this subtlety by introducing a so-called \emph{complexity
length scale} $\mathrm{L}_{\mathrm{C}}(t_{A}$, $t_{B})\geq s(t_{A}$, $t_{B})$
for any quantum evolution specified by a length $s$ of the path connecting an
initial and a final quantum state $\left\vert \psi_{A}\left(  \theta
_{A}\text{, }\varphi_{A}\right)  \right\rangle $ and $\left\vert \psi
_{B}\left(  \theta_{B}\text{, }\varphi_{B}\right)  \right\rangle $,
respectively. Finally, we define $\mathrm{L}_{\mathrm{C}}$ as%
\begin{equation}
\mathrm{L}_{\mathrm{C}}(t_{A}\text{, }t_{B})\overset{\text{def}}{=}%
\frac{s(t_{A}\text{, }t_{B})}{\sqrt{\frac{\overline{\mathrm{V}}(t_{A}\text{,
}t_{B})}{\mathrm{V}_{\max}(t_{A}\text{, }t_{B})}}}\text{,} \label{complexityL}%
\end{equation}
that is, using Eqs. (\ref{avgcomplexity}), (\ref{j3}), and (\ref{j4}),%
\begin{equation}
\mathrm{L}_{\mathrm{C}}\left(  t_{A}\text{, }t_{B}\right)  \overset
{\text{def}}{=}\left(  \int_{t_{A}}^{t_{B}}2\frac{\Delta E\left(  t\right)
}{\hslash}dt\right)  \left(  \frac{\frac{1}{t_{B}-t_{A}}\int_{t_{A}}^{t_{B}%
}\left\vert \int\int_{\mathcal{D}_{\text{\textrm{accessed}}}\left(
\theta\text{, }\varphi\right)  }\sqrt{g_{\mathrm{FS}}\left(  \theta\text{,
}\varphi\right)  }d\theta d\varphi\right\vert dt}{\left\vert \int
\int_{\mathcal{D}_{\text{\textrm{accessible}}}\left(  \theta\text{, }%
\varphi\right)  }\sqrt{g_{\mathrm{FS}}\left(  \theta\text{, }\varphi\right)
}d\theta d\varphi\right\vert }\right)  ^{-1/2}\text{.} \label{QCLS}%
\end{equation}
The quantity $\mathrm{L}_{\mathrm{C}}\left(  t_{A}\text{, }t_{B}\right)  $ in
Eq. (\ref{QCLS}) is the proposed quantum complexity length scale for a quantum
evolution that happens in a finite time interval $t_{B}-t_{A}$. From a
dimensional analysis standpoint, Eq. (\ref{complexityL}) can be seen as
emerging from the educated guess, $s^{2}:\overline{\mathrm{V}}=\mathrm{L}%
_{\mathrm{C}}^{2}:\mathrm{V}_{\max}$. Although we have been using the word
\textquotedblleft volume\textquotedblright\ in this paper, these parametric
regions in Eqs. (\ref{j5B}) and (\ref{j5}) are actually two-dimensional.
Loosely speaking, our intuitive perspective behind our proposed complexity
quantum length scale definition in Eq. (\ref{complexityL}) can be summarized
as follows. For a given pair of initial and final quantum states on the Bloch
sphere, describing a shorter path that connects these two states with an
accessed parameter space that is a big fraction of the locally accessible
parameter space is less complex than describing a longer path with an accessed
parameter space that is only a small fraction of the locally accessible
parameter space. This length scale $\mathrm{L}_{\mathrm{C}}$ is meant to
capture the following intuitive point: A longer path that occupies a smaller
fraction of the maximum volume of the accessible parametric region is more
complex than a shorter path that occupies as larger fraction of the available
parametric region. In principle, a more efficient quantum evolution need not
to be necessarily a less complex quantum evolution (i.e., $s_{1}\leq s_{2}$
does not necessarily imply $\mathrm{L}_{\mathrm{C}_{1}}\leq\mathrm{L}%
_{\mathrm{C}_{2}}$). Indeed, a less complex quantum evolution is not
necessarily the one that goes through a smaller average accessed volume in a
shorter time. Rather, it is the one for which the average accessed volume is
as close as possible to the maximum accessible volume. Our proposed complexity
length scale takes into account both effects, i.e. the length of the path and
the ratio between accessed and accessible volumes (or, in other words, the
complexity of the evolution). Indeed, using Eqs. (\ref{QCD}) and
(\ref{complexityL}), an alternative representation of $\mathrm{L}_{\mathrm{C}%
}$ is%
\begin{equation}
\mathrm{L}_{\mathrm{C}}(t_{A}\text{, }t_{B})=\frac{s(t_{A}\text{, }t_{B}%
)}{\sqrt{1-\mathrm{C}\left(  t_{A}\text{, }t_{B}\right)  }}\text{.}
\label{lisa}%
\end{equation}
From Eq. (\ref{lisa}), we see that $\mathrm{L}_{\mathrm{C}}(t_{A}$,
$t_{B})\geq s(t_{A}$, $t_{B})$ since $0\leq\mathrm{C}\left(  t_{A}\text{,
}t_{B}\right)  \leq1$ by construction. Furthermore, it is interesting to note
that\textbf{ }$s_{2}(t_{A}$, $t_{B})\geq s_{1}(t_{A}$, $t_{B})$ does not imply
necessarily that $\mathrm{L}_{\mathrm{C}_{2}}(t_{A}$, $t_{B})\geq
\mathrm{L}_{\mathrm{C}_{1}}(t_{A}$, $t_{B})$. Indeed, if $\mathrm{C}%
_{2}\left(  t_{A}\text{, }t_{B}\right)  $ is sufficiently smaller than
$\mathrm{C}_{1}\left(  t_{A}\text{, }t_{B}\right)  $ and, in addition,
$s_{2}(t_{A}$, $t_{B})$ is not too much greater than $s_{1}(t_{A}$, $t_{B})$,
we can still have $\mathrm{L}_{\mathrm{C}_{2}}(t_{A}$, $t_{B})\leq
\mathrm{L}_{\mathrm{C}_{1}}(t_{A}$, $t_{B})$. Thus, the complexity length
scale of quantum evolutions yielding longer paths can be smaller than the
complexity length scale of evolutions leading to shorter paths, provided that
their corresponding accessed volumes are as close as possible to their
accessible volumes. (i.e., $\mathrm{C}\left(  t_{A}\text{, }t_{B}\right)  $ as
small as possible).

We are now ready to apply our proposed measure of complexity $\mathrm{C}%
(t_{A}$, $t_{B})$ in Eq. (\ref{QCD}) and quantum complexity length scale
$\mathrm{L}_{\mathrm{C}}(t_{A}$, $t_{B})$ in Eq. (\ref{complexityL})\textbf{
}to actual physical evolutions for a two-level quantum system.\begin{table}[t]
\centering
\begin{tabular}
[c]{c|c|c|c|c|c}\hline\hline
$\alpha$ & $\overline{\mathrm{V}}\left(  \alpha\right)  $ & \textrm{V}$_{\max
}(\alpha)$ & \textrm{C}$\left(  \alpha\right)  $ & \textrm{L}$_{\mathrm{C}%
}(\alpha)$ & $\pi-\alpha$\\\hline
$0$ & $0.1917$ & $0.2777$ & $0.3096$ & $2.6735$ & $\pi$\\\hline
$\frac{1}{16}\pi$ & $0.1536$ & $0.2243$ & $0.3152$ & $2.3996$ & $\frac{15}%
{16}\pi$\\\hline
$\frac{1}{8}\pi$ & $0.1064$ & $0.1765$ & $0.3973$ & $2.3509$ & $\frac{7}{8}%
\pi$\\\hline
$\frac{3}{16}\pi$ & $0.0508$ & $0.1358$ & $0.6259$ & $2.8135$ & $\frac{13}%
{16}\pi$\\\hline
$\frac{1}{4}\pi$ & $0.0333$ & $0.1016$ & $0.6719$ & $2.8888$ & $\frac{3}{4}%
\pi$\\\hline
$\frac{5}{16}\pi$ & $0.0238$ & $0.0723$ & $0.6710$ & $2.8128$ & $\frac{11}%
{16}\pi$\\\hline
$\frac{3}{8}\pi$ & $0.0153$ & $0.0465$ & $0.6706$ & $2.7674$ & $\frac{5}{8}%
\pi$\\\hline
$\frac{7}{16}\pi$ & $0.0075$ & $0.0228$ & $0.6705$ & $2.7439$ & $\frac{9}%
{16}\pi$\\\hline
$\frac{1}{2}\pi$ & $0.3927$ & $0.7854$ & $0.5$ & $2.2214$ & $\frac{1}{2}\pi
$\\\hline
\end{tabular}
\caption{Numerical estimates of the accessed volume $\overline{\mathrm{V}%
}\left(  \alpha\right)  $, the accessible volume \textrm{V}$_{\max}(\alpha)$,
the complexity \textrm{C}$\left(  \alpha\right)  $, and the complexity length
scale \textrm{L}$_{\mathrm{C}}\left(  \alpha\right)  $ for $\alpha\in\left[
0\text{, }\pi\right]  $. The interval $\left[  0\text{, }\pi\right]  $ was
partitioned into sixteen sub-intervals of equal size $\pi/16$. Note that
$\overline{\mathrm{V}}\left(  \alpha\right)  $ and \textrm{V}$_{\max}(\alpha)$
assume the same numerical values for the angles $\alpha$ and the corresponding
supplementary angles $\pi-\alpha$. This symmetry is expected on physical
grounds.}%
\end{table}

\section{Complexity of Hamiltonian evolutions}

In this section, we numerically estimate the complexity measure $\mathrm{C}%
(t_{A}$, $t_{B})$ in Eq. (\ref{QCD}) and the complexity length scale
$\mathrm{L}_{\mathrm{C}}(t_{A}$, $t_{B})$ in Eq. (\ref{complexityL}) for
$\mathrm{H}_{\mathrm{opt}}$ in Eq. (\ref{Hopt}) and several members of the
family of Hamiltonians $\mathrm{H}_{\mathrm{sub}\text{\textrm{-}}\mathrm{opt}%
}\left(  \alpha\right)  $ in Eq. (\ref{Hsopt}) with $0\leq\alpha\leq\pi
$.\begin{figure}[t]
\centering
\includegraphics[width=0.4\textwidth] {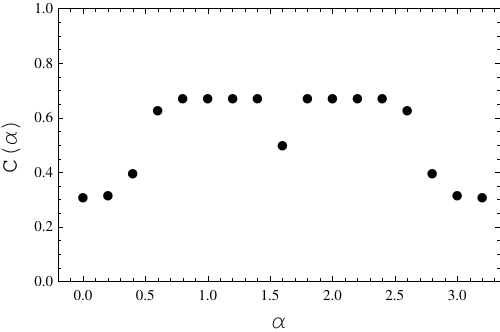}\caption{Scatter plot of the
complexity \textrm{C}$\left(  \alpha\right)  $ versus $\alpha$, with
$\alpha\in\left[  0\text{, }\pi\right]  $. The numerical estimates for
\textrm{C}$\left(  \alpha\right)  $ appear in Table I. For values of $\alpha$
in the interval centered at $\pi/2$ and radius $\pi/4$, i.e. in $\left[
\pi/4\text{, }(3/4)\pi\right]  $, the complexity \textrm{C}$\left(
\alpha\right)  $ assumes its local minimum value of $0.5$ at $\alpha=\pi/2$.
However, the complexity of Hamiltonian evolutions with $\alpha$ sufficiently
far from $\pi/2$ can assume values smaller than $0.5$. Recalling that a
complex Hamiltonian model is one with a small fraction $\overline{\mathrm{V}%
}\left(  \alpha\right)  /\mathrm{V}_{\max}\left(  \alpha\right)  $, this
behavior can be ascribed to the fact that these Hamiltonians \textrm{H}%
$_{\mathrm{sub}\text{-}\mathrm{opt}}\left(  \alpha\right)  $ yield longer
paths connecting initial and final states and tend to occupy a large fraction
of the accessible volume.}%
\end{figure}

In what follows, we assume for illustrative purposes to consider quantum
evolutions from the initial state $\left\vert A\right\rangle \overset
{\text{def}}{=}\left[  \left\vert 0\right\rangle +\left\vert 1\right\rangle
\right]  /\sqrt{2}$ with corresponding Bloch vector $\hat{a}\overset
{\text{def}}{=}(1$, $0$, $0)$ to the final state $\left\vert B\right\rangle
\overset{\text{def}}{=}\left[  \left\vert 0\right\rangle +i\left\vert
1\right\rangle \right]  /\sqrt{2}$ with corresponding Bloch vector $\hat
{b}\overset{\text{def}}{=}(0$, $1$, $0)$. Note that $\hat{a}\cdot\hat{b}%
=\cos\left(  \theta_{AB}\right)  =0$ since $\theta_{AB}=\pi/2$. Using the
optimal Hamiltonian $\mathrm{H}_{\mathrm{opt}}$ in Eq. (\ref{Hopt}), it turns
out that an intermediate state at time $t\in\left[  t_{A}\text{, }%
t_{B}\right]  $ is physically equivalent to the state $\left\vert
\psi(t)\right\rangle =(1/\sqrt{2})\left\vert 0\right\rangle +(1/\sqrt
{2})e^{2i\omega t}\left\vert 1\right\rangle $ (with $\omega\overset
{\text{def}}{=}E/\hslash$). Therefore, the spherical angles become
$\theta(t)=\pi/2$ and $\varphi(t)=2\omega t$. Furthermore, as a consistency
check, we point out that $t_{\mathrm{opt}}=\left[  \hslash/(2E)\right]
\theta_{AB}=\pi/(4\omega)$ and $\left\vert \psi(t_{\mathrm{opt}})\right\rangle
=\left[  \left\vert 0\right\rangle +i\left\vert 1\right\rangle \right]
/\sqrt{2}=\left\vert B\right\rangle $. In the sub-optimal scenario, using the
sub-optimal Hamiltonian in Eq. (\ref{Hsopt}) and standard quantum-mechanical
rules, the unitary evolution operator becomes%
\begin{equation}
U_{\mathrm{sub}\text{-\textrm{optimal}}}(t)\overset{\text{def}}{=}\left(
\begin{array}
[c]{cc}%
\cos(\omega t)-i\sin(\alpha)\sin(\omega t) & -\frac{\cos(\alpha)}{\sqrt{2}%
}(1+i)\sin(\omega t)\\
\frac{\cos(\alpha)}{\sqrt{2}}(1-i)\sin(\omega t) & \cos(\omega t)+i\sin
(\alpha)\sin(\omega t)
\end{array}
\right)  \text{.}\label{unitary}%
\end{equation}
Applying $U_{\mathrm{sub}\text{-\textrm{optimal}}}(t)$ to the initial state
$\left\vert A\right\rangle $, the intermediate state $\left\vert \psi\left(
t\right)  \right\rangle =U_{\mathrm{sub}\text{-\textrm{optimal}}}(t)\left\vert
A\right\rangle $ can be recast as $\left\vert \psi\left(  t\right)
\right\rangle =c_{0}(t)\left\vert 0\right\rangle +c_{1}(t)\left\vert
1\right\rangle $. After some algebra, the complex probability amplitudes
$c_{0}(t)=c_{0}\left(  t\text{; }\alpha\text{, }\omega\right)  $ and
$c_{1}(t)=c_{1}\left(  t\text{; }\alpha\text{, }\omega\right)  $ can be
expressed as%
\begin{equation}
c_{0}\left(  t\text{; }\alpha\text{, }\omega\right)  \overset{\text{def}}%
{=}\left(  \frac{\cos(\omega t)}{\sqrt{2}}-\frac{\cos(\alpha)\sin(\omega
t)}{2}\right)  +i\left(  -\frac{\cos(\alpha)\sin(\omega t)}{2}-\frac
{\sin(\alpha)\sin\left(  \omega t\right)  }{\sqrt{2}}\right)  \text{,}%
\label{amplitude1}%
\end{equation}
and,%
\begin{equation}
c_{1}\left(  t\text{; }\alpha\text{, }\omega\right)  \overset{\text{def}}%
{=}\left(  \frac{\cos(\omega t)}{\sqrt{2}}+\frac{\cos(\alpha)\sin(\omega
t)}{2}\right)  +i\left(  -\frac{\cos(\alpha)\sin(\omega t)}{2}+\frac
{\sin(\alpha)\sin\left(  \omega t\right)  }{\sqrt{2}}\right)  \text{.}%
\label{amplitude2}%
\end{equation}
As a side remark, we note that $\left\vert c_{0}(t)\right\vert ^{2}+\left\vert
c_{1}(t)\right\vert ^{2}=1$. Moreover, $c_{0}(t)\rightarrow(1/\sqrt
{2})e^{-i\omega t}$ and $c_{1}(t)\rightarrow(1/\sqrt{2})e^{i\omega t}$ as
$\alpha$ approaches $\pi/2$, as expected. Using Eqs. (\ref{teta}), (\ref{fi}),
(\ref{amplitude1}), and (\ref{amplitude2}), we get%
\begin{equation}
\theta\left(  t\text{; }\alpha\text{, }\omega\right)  \overset{\text{def}}%
{=}2\arctan\left\{  \sqrt{\frac{\frac{1}{2}-\frac{\sin(2\alpha)}{2\sqrt{2}%
}\sin^{2}(\omega t)+\frac{\cos(\alpha)}{2\sqrt{2}}\sin(2\omega t)}{\frac{1}%
{2}+\frac{\sin(2\alpha)}{2\sqrt{2}}\sin^{2}(\omega t)-\frac{\cos(\alpha
)}{2\sqrt{2}}\sin(2\omega t)}}\right\}  \text{,}\label{yoteta}%
\end{equation}
and,%
\begin{equation}
\varphi\left(  t\text{; }\alpha\text{, }\omega\right)  \overset{\text{def}}%
{=}\arctan\left[  \frac{\frac{\sin(\alpha)\sin(\omega t)}{\sqrt{2}}-\frac
{\cos(\alpha)\sin(\omega t)}{2}}{\frac{\cos(\alpha)\sin(\omega t)}{2}%
+\frac{\cos(\omega t)}{\sqrt{2}}}\right]  -\arctan\left[  \frac{\frac
{\cos(\alpha)\sin(\omega t)}{2}+\frac{\sin(\alpha)\sin(\omega t)}{\sqrt{2}}%
}{\frac{\cos(\alpha)\sin(\omega t)}{2}-\frac{\cos(\omega t)}{\sqrt{2}}%
}\right]  \text{.}\label{yofi}%
\end{equation}
As a first consistency check, note that we correctly get $\theta\left(
0\text{; }\alpha\text{, }\omega\right)  =\pi/2$ and $\varphi\left(  0\text{;
}\alpha\text{, }\omega\right)  =0$. However, we must keep in mind that the
expression for $\varphi\left(  t\right)  $ in Eq. (\ref{yofi}) can become more
complicated. As we previously mentioned, this is caused by the fact that the
phase $\arg\left(  z\right)  $ of a complex number $%
%TCIMACRO{\U{2102} }%
%BeginExpansion
\mathbb{C}
%EndExpansion
\ni z\overset{\text{def}}{=}x+iy=\left\vert z\right\vert e^{i\arg(z)}$ must be
generally expressed by means of the $2$-\textrm{argument arctangent} function
\textrm{atan}$2$ as $\arg(z)=$\textrm{atan}$2(y$, $x)$. When $x>0$,
\textrm{atan}$2(y$, $x)$ reduces to $\arctan\left(  y/x\right)  $. This point
in especially important in the numerical estimations of the complexity. For
more details, we refer to Appendix A (Complexity Symmetries) and Appendix B
(Complexity Calculations). Then, using Eqs. (\ref{yoteta}), (\ref{yofi}), and
(\ref{local-complexity}), the time-dependent volume $V\left(  t\right)  $ in
Eq. (\ref{local-complexity}) reduces to%
\begin{equation}
V\left(  t\text{; }\alpha\text{, }\omega\right)  =\frac{1}{4}\left\vert
\varphi(t\text{; }\alpha\text{, }\omega)\cos\left[  \theta\left(  t\text{;
}\alpha\text{, }\omega\right)  \right]  \right\vert \text{,}\label{vt}%
\end{equation}
given that $\varphi(0$; $\alpha$, $\omega)=0$ and $\theta\left(  0\text{;
}\alpha\text{, }\omega\right)  =\pi/2$. From Eq. (\ref{vt}), one can calculate
the accessed volume $\overline{\mathrm{V}}\left(  \alpha\text{, }%
\omega\right)  $ in Eq. (\ref{avgcomplexity}) and obtain%
\begin{equation}
\overline{\mathrm{V}}\left(  \alpha\text{, }\omega\right)  \overset
{\text{def}}{=}\frac{1}{t_{\mathrm{sub}\text{-\textrm{opt}}}}\int
_{0}^{t_{\mathrm{sub}\text{-\textrm{opt}}}}V\left(  t\text{; }\alpha\text{,
}\omega\right)  dt\text{,}\label{eros}%
\end{equation}
where $t_{\mathrm{sub}\text{-\textrm{opt}}}=t_{\mathrm{sub}\text{-\textrm{opt}%
}}\left(  \alpha\text{, }\omega\right)  $ is given in Eq. (\ref{local}). Using
Eqs. (\ref{teta}), (\ref{fi}), and (\ref{local}), one can also numerically
calculate the lower and upper bounds in Eq. (\ref{minmax}). Then, one uses
these bounds to estimate the accessible volume \textrm{V}$_{\max}(\alpha$,
$\omega)$ in Eq. (\ref{j4}). Finally, having an estimate of $\overline
{\mathrm{V}}\left(  \alpha\text{, }\omega\right)  $, \textrm{V}$_{\max}%
(\alpha$, $\omega)$, and $s\left(  \alpha\right)  $ in Eq. (\ref{yoyo}), one
can arrive at the numerical estimates for the complexity \textrm{C}$\left(
\alpha\text{, }\omega\right)  \overset{\text{def}}{=}\left[  \mathrm{V}_{\max
}(\alpha,\omega)-\overline{\mathrm{V}}\left(  \alpha\text{, }\omega\right)
\right]  /$\textrm{V}$_{\max}(\alpha$, $\omega)$ in Eq. (\ref{QCD}) and the
complexity length scale \textrm{L}$_{\mathrm{C}}\left(  \alpha\text{, }%
\omega\right)  \overset{\text{def}}{=}s\left(  \alpha\right)  /\sqrt
{1-\mathrm{C}\left(  \alpha\text{, }\omega\right)  }$ in Eq.
(\ref{complexityL}). In Table I, we set $\omega=1$ and report the numerical
estimates of the accessed volume $\overline{\mathrm{V}}\left(  \alpha\right)
$, the accessible volume \textrm{V}$_{\max}(\alpha)$, the complexity
\textrm{C}$\left(  \alpha\right)  $, and the complexity length scale
\textrm{L}$_{\mathrm{C}}\left(  \alpha\right)  $ for $0\leq\alpha\leq\pi$.
Furthermore, we keep $\omega=1$ and illustrate in Figs.\textbf{ }$4$\textbf{
}and\textbf{ }$5$\textbf{ }the scatter plots of the complexity \textrm{C}%
$\left(  \alpha\right)  $ and the complexity length scale \textrm{L}%
$_{\mathrm{C}}\left(  \alpha\right)  $ versus $\alpha$, respectively. Finally,
in Table II, we maintain $\omega=1$ and display the numerical estimates of the
geodesic efficiency $\mathrm{\eta}_{\mathrm{GE}}\left(  \alpha\right)  $ in
Eq. (\ref{SOGE}), the speed efficiency $\eta_{\mathrm{SE}}(\alpha)$ in Eq.
(\ref{SOGE1}), the curvature coefficient $\kappa_{\mathrm{AC}}^{2}\left(
\alpha\right)  $ in Eq. (\ref{SOGE2}), the complexity \textrm{C}$\left(
\alpha\right)  $ in Eq. (\ref{QCD}), and the complexity length scale
\textrm{L}$_{\mathrm{C}}\left(  \alpha\right)  $ in Eq. (\ref{complexityL})
for $0\leq\alpha\leq\pi$. Interestingly, we point out that when inspecting the
numerical estimates in Table II, one can notice that the curvature coefficient
$\kappa_{\mathrm{AC}}^{2}\left(  \alpha\right)  $ versus the efficiencies
$\eta_{\mathrm{GE}}\left(  \alpha\right)  $ or $\eta_{\mathrm{SE}}\left(
\alpha\right)  $) exhibits a monotonically decreasing behavior. Instead, the
complexity \textrm{C}$\left(  \alpha\right)  $ and the complexity length scale
\textrm{L}$_{\mathrm{C}}\left(  \alpha\right)  $ display a non-monotonic
behavior when viewed as functions of $\eta_{\mathrm{GE}}\left(  \alpha\right)
$ or $\eta_{\mathrm{SE}}\left(  \alpha\right)  $. Additional remarks on these
intriguing trends appear in the next section\textbf{. }It is also important to
point out that for the class of Hamiltonian evolutions analyzed here, the
accessed volume $\overline{\mathrm{V}}\left(  \alpha\text{, }\omega\right)  $
and the accessible volume \textrm{V}$_{\max}\left(  \alpha\text{, }%
\omega\right)  $ do not depend on $\omega\overset{\text{def}}{=}E/\hslash$.
Furthermore, it happens that $\overline{\mathrm{V}}\left(  \alpha\right)
=\overline{\mathrm{V}}\left(  \pi-\alpha\right)  $ and \textrm{V}$_{\max
}\left(  \alpha\right)  =$\textrm{V}$_{\max}\left(  \pi-\alpha\right)  $.
Thus, the complexity \textrm{C}$\left(  \alpha\right)  $ and the complexity
length scale \textrm{L}$_{\mathrm{C}}\left(  \alpha\right)  $ possess
identical values for supplementary angles $\alpha$ and $\pi-\alpha$ for any
choice of the value for the frequency $\omega$ (or, alternatively, for any
finite energy value $E$). These symmetries are discussed in Appendix A.
Instead, technical details on the numerical estimates of \textrm{C}$\left(
\alpha\right)  $ can be found in Appendix B.

We are now ready for our summary of results and final
remarks.\begin{figure}[t]
\centering
\includegraphics[width=0.4\textwidth] {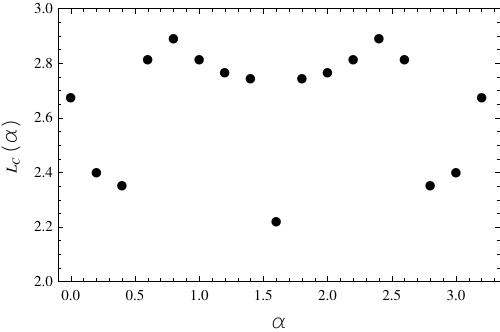}\caption{Scatter plot of the
complexity length scale \textrm{L}$_{\mathrm{C}}\left(  \alpha\right)  $
versus $\alpha$, with $\alpha\in\left[  0\text{, }\pi\right]  $. The numerical
estimates for \textrm{C}$\left(  \alpha\right)  $ appear in Table I. The
minimum value of \textrm{L}$_{\mathrm{C}}\left(  \alpha\right)  $ is obtained
at $\alpha=\pi/2\simeq1.57$. Focusing on the interval with $\alpha\in\left[
0\text{, }\pi/2\right]  $, we note that \textrm{L}$_{\mathrm{C}}\left(
\alpha\right)  $ exhibits a non-monotonic behavior. In particular, the
complexity length scale of Hamiltonian evolutions specified by values of
$\alpha$ sufficiently far from $\pi/2$ can be smaller than that of evolutions
characterized by values of $\alpha$ close to $\pi/2$. This is due to the fact
that although Hamiltonians \textrm{H}$_{\mathrm{sub}\text{-}\mathrm{opt}%
}\left(  \alpha\right)  $ specified by values $\alpha_{1}$ of $\alpha$
sufficiently far from $\pi/2$ yield paths connecting initial and final states
that are longer than the ones generated by evolutions characterized by values
$\alpha_{2}$ of $\alpha$ close to $\pi/2$, they also occupy a larger fraction
of the accessible volume than that of evolutions with $\alpha$ close to
$\pi/2$. In other words, although $s_{1}>s_{2}$, $\left(  \overline
{\mathrm{V}}_{1}/\mathrm{V}_{\max\text{-}1}\right)  ^{1/2}\gg(\overline
{\mathrm{V}}_{2}/\mathrm{V}_{\max\text{-}2})^{1/2}$. Thus, being
\textrm{L}$_{\mathrm{C}}\left(  \alpha\right)  \overset{\text{def}}{=}%
s\cdot\left(  \overline{\mathrm{V}}/\mathrm{V}_{\max}\right)  ^{-1/2}$,
\textrm{L}$_{\mathrm{C}}\left(  \alpha_{1}\right)  $ becomes smaller than
\textrm{L}$_{\mathrm{C}}\left(  \alpha_{2}\right)  $.}%
\end{figure}\begin{table}[t]
\centering
\begin{tabular}
[c]{c|c|c|c|c|c|c}\hline\hline
$\alpha$ & $\eta_{\mathrm{GE}}\left(  \alpha\right)  $ & $\eta_{\mathrm{SE}%
}\left(  \alpha\right)  $ & $\kappa_{\mathrm{AC}}^{2}\left(  \alpha\right)  $
& \textrm{C}$\left(  \alpha\right)  $ & \textrm{L}$_{\mathrm{C}}\left(
\alpha\right)  $ & $\pi-\alpha$\\\hline
$0$ & $0.7071$ & $0.7071$ & $4$ & $0.3096$ & $2.6735$ & $\pi$\\\hline
$\frac{1}{16}\pi$ & $0.7911$ & $0.7204$ & $3.7067$ & $0.3152$ & $2.3996$ &
$\frac{15}{16}\pi$\\\hline
$\frac{1}{8}\pi$ & $0.8607$ & $0.7571$ & $2.9781$ & $0.3973$ & $2.3509$ &
$\frac{7}{8}\pi$\\\hline
$\frac{3}{16}\pi$ & $0.9128$ & $0.8089$ & $2.1131$ & $0.6259$ & $2.8135$ &
$\frac{13}{16}\pi$\\\hline
$\frac{1}{4}\pi$ & $0.9493$ & $0.8660$ & $1.3333$ & $0.6719$ & $2.8888$ &
$\frac{3}{4}\pi$\\\hline
$\frac{5}{16}\pi$ & $0.9737$ & $0.9196$ & $0.7298$ & $0.6710$ & $2.8128$ &
$\frac{11}{16}\pi$\\\hline
$\frac{3}{8}\pi$ & $0.9890$ & $0.9627$ & $0.3160$ & $0.6706$ & $2.7674$ &
$\frac{5}{8}\pi$\\\hline
$\frac{7}{16}\pi$ & $0.9973$ & $0.9904$ & $0.0776$ & $0.6705$ & $2.7439$ &
$\frac{9}{16}\pi$\\\hline
$\frac{1}{2}\pi$ & $1$ & $1$ & $0$ & $0.5$ & $2.2214$ & $\frac{1}{2}\pi
$\\\hline
\end{tabular}
\caption{Numerical estimates of the geodesic efficiency $\mathrm{\eta
}_{\mathrm{GE}}\left(  \alpha\right)  $, the speed efficiency $\eta
_{\mathrm{SE}}(\alpha)$, the curvature coefficient $\kappa_{\mathrm{AC}}%
^{2}\left(  \alpha\right)  $, the complexity \textrm{C}$\left(  \alpha\right)
$, and the complexity length scale \textrm{L}$_{\mathrm{C}}\left(
\alpha\right)  $ for $\alpha\in\left[  0\text{, }\pi\right]  $. The interval
$\left[  0\text{, }\pi\right]  $ was partitioned into sixteen sub-intervals of
equal size $\pi/16$. Note that all quantities exhibit the same numerical
values for the angles $\alpha$ and the corresponding supplementary angles
$\pi-\alpha$. This symmetry is expected on physical grounds.}%
\end{table}

\section{Concluding Remarks}

In this paper, we studied the complexity of both time optimal (Eq.
(\ref{Hopt})) and time sub-optimal (Eq. (\ref{Hsopt})) quantum Hamiltonian
evolutions connecting arbitrary source and target states on the Bloch sphere
equipped with the Fubini-Study metric. This study was conducted in several
steps. Initially, we characterized each unitary Schr\"{o}dinger quantum
evolution by quantifying its path length (Eq. (\ref{yoyo})), geodesic
efficiency (Eq. (\ref{SOGE})), speed efficiency (Eq. (\ref{SOGE1})), and
curvature coefficient (Eq. (\ref{SOGE2})). These quantities were used to
describe the dynamical trajectory connecting the source state to the target
state (Fig. $2$). Next, we went from a classical probabilistic framework,
where the complexity of entropic motion on curved statistical manifolds can be
described using the information geometric complexity in Eq. (\ref{IGC}), to a
deterministic quantum setting. In this context, we proposed a definition of
quantum evolution complexity (Eq. (\ref{QCD}) and Fig.\textbf{ }$3$) and
introduced the concept of a quantum complexity length scale (Eq.
(\ref{complexityL})). We discussed the physical significance of these
quantities in terms of the partial (i.e., accessed volume in Eq.
(\ref{avgcomplexity})) and total (i.e., accessible volume in Eq. (\ref{j4}))
parametric volumes of the regions on the Bloch sphere that specify the quantum
mechanical evolution from source to target states. Furthermore, we numerically
estimated the complexity measure (Table I and Fig\textbf{. }$4$) and the
complexity length scale (Table I and Fig\textbf{. }$5$) for several quantum
evolutions specified by Hamiltonians in Eq. (\ref{Hsopt}) for different values
of the parameter $\alpha$ that allows to transition from optimal to
sub-optimal quantum evolution scenarios. Finally, we compared the behavior of
our proposed complexity measures with that of the geodesic efficiency, the
speed efficiency, and the curvature coefficient by means of numerical
estimates as displayed in Table II. Our work is one of a kind, relating
concepts such as complexity, curvature, and efficiency within a single
theoretical platform. The main take-home messages are:

\begin{enumerate}
\item[{[i]}] Complexity is more than length. Specifically, quantum evolutions
generated by Hamiltonians \textrm{H}$_{\mathrm{sub}\text{-}\mathrm{opt}%
}\left(  \alpha\right)  $ in Eq. (\ref{Hsopt}) that connect initial and final
states along a shorter path on the Bloch sphere do not need to be necessarily
less complex than quantum evolutions connecting the same initial and final
states along longer paths. In other words, $\eta_{\mathrm{GE}}\left(
\alpha_{1}\right)  <\eta_{\mathrm{GE}}\left(  \alpha_{2}\right)  $ does not
generally imply that \textrm{C}$\left(  \alpha_{1}\right)  >$\textrm{C}%
$\left(  \alpha_{2}\right)  $.

\item[{[ii]}] Wasting energy resources does not lead automatically to more
complex quantum evolutions. Stated otherwise, $\eta_{\mathrm{SE}}\left(
\alpha_{1}\right)  <\eta_{\mathrm{SE}}\left(  \alpha_{2}\right)  $ does not
generally suggest that \textrm{C}$\left(  \alpha_{1}\right)  >$\textrm{C}%
$\left(  \alpha_{2}\right)  $.

\item[{[iii]}] Quantum evolutions yielding longer paths that are sufficiently
bent can be less complex than quantum evolutions specified by shorter paths
that are less bent. Put differently, $\mathrm{\kappa}_{\mathrm{AC}}^{2}\left(
\alpha_{1}\right)  >\mathrm{\kappa}_{\mathrm{AC}}^{2}\left(  \alpha
_{2}\right)  $ does not generally signify that \textrm{C}$\left(  \alpha
_{1}\right)  >$\textrm{C}$\left(  \alpha_{2}\right)  $.

\item[{[iv]}] If one wishes to evolve on the Bloch sphere in the shortest
possible time, with maximum speed and no waste of energy resources, the
Hamiltonian \textrm{H}$_{\mathrm{opt}}=$\textrm{H}$_{\mathrm{sub}%
\text{-}\mathrm{opt}}\left(  \pi/2\right)  $ remains the best choice, even in
terms of our proposed complexity length scale \textrm{L}$_{\mathrm{C}}\left(
\alpha\right)  $. However, if time is not an issue and wasting large amounts
of energy resources is tolerated, it is possible to evolve along paths
characterized by complexity values smaller than those that specify the
shortest possible paths generated by \textrm{H}$_{\mathrm{opt}}$. In other
words, it can be \textrm{C}$\left(  \alpha_{1}\right)  <$\textrm{C}$\left(
\alpha_{2}\right)  $ despite the fact that $\eta_{\mathrm{GE}}\left(
\alpha_{1}\right)  <\eta_{\mathrm{GE}}\left(  \alpha_{2}\right)  $ and
$\eta_{\mathrm{SE}}\left(  \alpha_{1}\right)  <\eta_{\mathrm{SE}}\left(
\alpha_{2}\right)  $. This scenario happens when $\overline{\mathrm{V}}\left(
\alpha_{1}\right)  /\mathrm{V}_{\max}\left(  \alpha_{1}\right)  \gg
\overline{\mathrm{V}}\left(  \alpha_{2}\right)  /\mathrm{V}_{\max}\left(
\alpha_{2}\right)  $. This set of inequalities is in agreement with the fact
that complexity is more than length. This latter statement, in turn, is
well-captured by our proposed complexity length scale \textrm{L}$_{\mathrm{C}%
}\left(  \alpha\right)  $.

\item[{[v]}] There is no free lunch: If one chooses to move fast and with no
waste of energy resources, one must accept some non-minimal level of
complexity. If one wishes to minimize the level of complexity, instead, one
must give up on speed and on saving energy resources. In any case, our
analysis suggests that it is better to pick one of the two extreme scenarios
and neglect intermediate cases specified by moderate levels of complexity that
do not require so much time and waste of energy resources.
\end{enumerate}

\medskip

The geometry of quantum state manifolds employed in our work, along with the
probabilistic structure underlying the concept of information geometric
complexity which is behind our newly proposed quantum complexity measure,
allowed us to present in a unifying manner links among quantities such as
curvature, efficiency, and complexity measures for quantum evolutions. This is
a nontrivial finding in itself. Firstly, focusing on the link between
complexity and curvature, we think it would be very helpful carrying out a
comparative investigation between our curvature coefficient and the four-index
Riemannian curvature introduced in Ref. \cite{hetenyi23}. Indeed, the
four-index Riemannian curvature employed in Ref. \cite{hetenyi23} was shown to
be in a one-to-one correspondence with the fourth cumulant (i.e., the
kurtosis) \ of a cumulant generating function characterizing the Riemannian
geometry of the state space of the quantum system being studied. In this
context, the Riemann curvature tensor would encode information on the
complexity of motion on the manifold of quantum states. Secondly, despite
requiring much more study, our quantitative discussion on the connection
between lengths and complexity measures defined in terms of volumes
originating from probability amplitudes is reminiscent of the rather
surprising connection between Krylov's and Nielsen's complexities reported in
Ref. \cite{craps24}.

\medskip

Our work has two obvious restrictions. First, we limited our applications to
quantum evolutions governed by stationary Hamiltonians for two-level quantum
systems. Second, we neglected any discussion on its applicability to the
characterization of the complexity of quantum-mechanical systems in mixed
quantum states. In principle, the extension of our work to time-varying
Hamiltonian evolutions is conceptually straightforward However, from a
computational standpoint, calculating the expression of probability amplitudes
requires precise analytical solutions to the time-dependent Schr\"{o}dinger
equation. This poses significant challenges, even in the context of two-level
quantum systems
\cite{landau32,zener32,rabi37,rabi54,barnes12,barnes13,messina14,grimaudo18,cafaroijqi,castanos,grimaudo23,elena20}%
. Furthermore, the shift from pure to mixed states presents numerous
challenges, both in computational and conceptual aspects. For example, it is
established that there exists an infinite variety of distinguishability
metrics for mixed quantum states \cite{karol06}. This variability results in
interpretations of significant geometric quantities, such as complexity and
volume of quantum states, that are dependent on the chosen metric.
Specifically, due to the non-uniqueness of these distinguishability distances,
it is essential to comprehend the physical significance of selecting a
particular metric, which remains a matter of considerable conceptual and
practical importance \cite{silva21,mera22,luongo24,chien24}. Given that these
above-mentioned complications are beyond the scope of our current work, we
will strive to tackle some aspects of these intricacies in forthcoming
scientific endeavors.

\medskip

Regardless of being limited to complexity investigations of geometric style
restricted to single-qubit evolutions like the one in Ref. \cite{marrone19},
it paves the way to several further investigations. First, we only compared
geodesic unwasteful evolutions with nongeodesic wasteful evolutions. However,
the spectrum of dynamical possibilities is much richer
\cite{campa19,campaioli19,xu24}. In particular, it would be interesting to
compare (in terms of complexity measures) also geodesic unwasteful paths with
geodesic wasteful paths or, alternatively, with nongeodesic unwasteful paths.
Second, we can extend the comparative analysis to nonstationary Hamiltonian
evolutions specified by physically significant time-dependent magnetic field
configurations \cite{messina14,grimaudo18,cafaroijqi,castanos,grimaudo23}.
Third, to investigate the connection between complexity and entanglement, we
can expand our approach to $d$-level quantum systems with $d>2$ in larger
finite-dimensional Hilbert spaces
\cite{jakob01,kimura03,krammer08,kurzy11,xie20,siewert21}. Lastly, we can
consider the possibility of offering a comparative analysis of the complexity
of quantum evolutions inside the Bloch sphere for quantum systems in mixed
quantum states \cite{hornedal22,nade24}.

\medskip

Regardless of its limitations, we think our work may inspire other scientists
to further investigate this research direction. We also believe that our
findings can lead to new extensions and applications in the future. We leave a
more detailed discussion on these potential advancements for forthcoming
scientific efforts.

\begin{acknowledgments}
C.C. thanks Emma Clements and Dominic Monaco for helpful comments and
discussions, respectively. Furthermore, the authors are grateful to anonymous
referees for constructive comments leading to an improved version of the
paper.\textbf{ }Any opinions, findings and conclusions or recommendations
expressed in this material are those of the author(s) and do not necessarily
reflect the views of their home Institutions.
\end{acknowledgments}

\pagebreak

\appendix

\section{Complexity symmetries}

In this Appendix, we demonstrate that for the class of Hamiltonian evolutions
considered in Eq. (\ref{Hsopt}), the accessed volume $\overline{\mathrm{V}%
}\left(  \alpha\text{, }\omega\right)  $ and the accessible volume
\textrm{V}$_{\max}\left(  \alpha\text{, }\omega\right)  $ do not depend on
$\omega\overset{\text{def}}{=}E/\hslash$. Furthermore, we show that
$\overline{\mathrm{V}}\left(  \alpha\right)  =\overline{\mathrm{V}}\left(
\pi-\alpha\right)  $ and \textrm{V}$_{\max}\left(  \alpha\right)  =$%
\textrm{V}$_{\max}\left(  \pi-\alpha\right)  $. Thus, the complexity
\textrm{C}$\left(  \alpha\right)  $ and the complexity length scale
\textrm{L}$_{\mathrm{C}}\left(  \alpha\right)  $ assume identical values for
supplementary angles $\alpha$ and $\pi-\alpha$ for any choice of the value for
the frequency $\omega$ (or, alternatively, for any finite energy value $E$).

To show that $\overline{\mathrm{V}}\left(  \alpha\text{, }\omega\right)  $
does not depend on $\omega$, we note that the time-dependent volume $V(t$;
$\alpha$, $\omega)$ in Eq. (\ref{vt}) is such that $V(t$; $\alpha$,
$\omega)=V(\omega t$; $\alpha)$. Therefore, $\overline{\mathrm{V}}\left(
\alpha\text{, }\omega\right)  $ can be recast as%
\begin{equation}
\overline{\mathrm{V}}\left(  \alpha\text{, }\omega\right)  =\frac{1}{t_{\ast}%
}\int_{0}^{t_{\ast}}V(\omega t\text{; }\alpha)dt\text{,} \label{eros1}%
\end{equation}
where $t_{\ast}\overset{\text{def}}{=}f\left(  \alpha\text{, }\theta
_{AB}\right)  /\omega$ and $f\left(  \alpha\text{, }\theta_{AB}\right)
\overset{\text{def}}{=}\omega t_{AB}\left(  \alpha\right)  $ with
$t_{AB}\left(  \alpha\right)  $ in Eq. (\ref{local}). Performing a change of
the integration variable $t$ in Eq. (\ref{eros1}) and setting $\tau
\overset{\text{def}}{=}\omega t$, $\overline{\mathrm{V}}\left(  \alpha\text{,
}\omega\right)  $ in Eq. (\ref{eros1}) reduces to%
\begin{equation}
\overline{\mathrm{V}}\left(  \alpha\text{, }\omega\right)  =\frac{1}{f\left(
\alpha\text{, }\theta_{AB}\right)  }\int_{0}^{f(\alpha\text{, }\theta_{AB}%
)}V(\tau\text{; }\alpha)d\tau\overset{\text{def}}{=}\overline{\mathrm{V}%
}\left(  \alpha\right)  \text{,} \label{eros2}%
\end{equation}
that is, $\overline{\mathrm{V}}\left(  \alpha\text{, }\omega\right)
=\overline{\mathrm{V}}\left(  \alpha\right)  $. Clearly, for different values
of $\theta_{AB}$, we generally get different values for $\overline{\mathrm{V}%
}\left(  \alpha\right)  $. In our applications, $\theta_{AB}$ equals $\pi/2$.
To understand that \textrm{V}$_{\max}\left(  \alpha\text{, }\omega\right)  $
does not depend on $\omega$, it is sufficient to note that optimizing
$\theta\left(  t\text{; }\alpha\text{, }\omega\right)  $ (or, analogously,
$\varphi\left(  t\text{; }\alpha\text{, }\omega\right)  $) over the time
interval $0\leq t\leq t_{\ast}\overset{\text{def}}{=}f\left(  \alpha\text{,
}\theta_{AB}\right)  /\omega$ is equivalent to optimizing $\theta\left(
\omega t\text{; }\alpha\right)  $ over $0\leq\omega t\leq f\left(
\alpha\text{, }\theta_{AB}\right)  $ (given the temporal behaviors of
$\theta\left(  t\text{; }\alpha\text{, }\omega\right)  $ and $\varphi\left(
t\text{; }\alpha\text{, }\omega\right)  $ in Eqs. (\ref{yoteta}) and
(\ref{yofi}), respectively). Therefore, putting $\tau\overset{\text{def}}%
{=}\omega t$, the optimization of $\theta\left(  t\text{; }\alpha\text{,
}\omega\right)  $ over $0\leq t\leq f\left(  \alpha\text{, }\theta
_{AB}\right)  /\omega$ is the same as the optimization of $\theta\left(
\tau\text{; }\alpha\right)  $ over $0\leq\tau\leq f\left(  \alpha\text{,
}\theta_{AB}\right)  $. Therefore, we conclude that \textrm{V}$_{\max}\left(
\alpha\text{, }\omega\right)  $ is independent from $\omega$. To be more
explicit, we remark that $\theta\left(  t\text{; }\alpha\text{, }\omega
_{1}\right)  $ and $\theta\left(  t\text{; }\alpha\text{, }\omega_{2}\right)
$ with $\omega_{1}\neq\omega_{2}$ assume the same minimum and maximum values.
However, they occur at different times $t_{1}\overset{\text{def}}{=}f\left(
\alpha\text{, }\theta_{AB}\right)  /\omega_{1}$ and $t_{2}\overset{\text{def}%
}{=}f\left(  \alpha\text{, }\theta_{AB}\right)  /\omega_{2}$, respectively.
Furthermore, to understand the reason why $\overline{\mathrm{V}}\left(
\alpha\right)  =\overline{\mathrm{V}}\left(  \pi-\alpha\right)  $, it suffices
observing that $f\left(  \alpha\text{, }\theta_{AB}\right)  $ in\ Eq.
(\ref{eros2}) equals $\omega t_{AB}\left(  \alpha\right)  $ with
$t_{AB}\left(  \alpha\right)  $ in Eq. (\ref{local}) and, thus, is such that
$f\left(  \alpha\text{, }\theta_{AB}\right)  =f\left(  \pi-\alpha\text{,
}\theta_{AB}\right)  $. This latter relation, in turn, is a consequence of the
fact that $t_{AB}\left(  \alpha\right)  =t_{AB}\left(  \pi-\alpha\right)  $
(given that $\sin\left(  \pi-\alpha\right)  =\sin\left(  \alpha\right)  $,
$\cos^{2}\left(  \pi-\alpha\right)  =\cos^{2}(\alpha)$, and $\sin^{2}%
(\pi-\alpha)=\sin^{2}(\alpha)$). Finally, to justify why \textrm{V}$_{\max
}\left(  \alpha\right)  =$\textrm{V}$_{\max}\left(  \pi-\alpha\right)  $, we
begin by noting that $\theta_{\min}\left(  \alpha\right)  =\pi/2$ and
$\theta_{\max}(\alpha)=\xi$ for some $\xi>\pi/2$. Then, geometric arguments
require $\theta_{\max}\left(  \alpha\right)  -\theta_{\min}\left(
\alpha\right)  =\theta_{\max}\left(  \pi-\alpha\right)  -\theta_{\min}\left(
\pi-\alpha\right)  $ and $\theta_{\max}\left(  \pi-\alpha\right)
=\theta_{\min}\left(  \alpha\right)  =\pi/2$. Therefore, we get $\theta_{\min
}\left(  \pi-\alpha\right)  =\pi-\xi$. Observe that for both the $\alpha$ and
$\pi-\alpha$ angles, the azimuthal angle belongs to the interval $\left[
0\text{, }\pi/2\right]  $ with $\varphi_{\min}=0$ and $\varphi_{\max}=\pi/2$.
Therefore, given these lower and upper bounds for the polar angles,
\textrm{V}$_{\max}\left(  \alpha\right)  =$\textrm{V}$_{\max}\left(
\pi-\alpha\right)  $ follows from the fact that%
\begin{equation}
\int_{\frac{\pi}{2}}^{\xi}\sin(\theta)d\theta=\int_{\pi-\xi}^{\frac{\pi}{2}%
}\sin(\theta)d\theta\text{.} \label{eros4}%
\end{equation}
With the presentation of the identity in Eq. (\ref{eros4}), we end our
discussion on some essential symmetry properties exhibited by the complexity
\textrm{C}$\left(  \alpha\right)  \overset{\text{def}}{=}$ $\left[
\mathrm{V}_{\max}\left(  \alpha\right)  -\overline{\mathrm{V}}\left(
\alpha\right)  \right]  /\mathrm{V}_{\max}\left(  \alpha\right)  $ and the
complexity length scale \textrm{L}$_{\mathrm{C}}\left(  \alpha\right)
\overset{\text{def}}{=}s\left(  \alpha\right)  /\sqrt{1-\mathrm{C}\left(
\alpha\right)  }$.

\section{Complexity calculations}

In this Appendix, we show some technical details on how to calculate the
complexity \textrm{C}$\left(  \alpha\right)  $ in Eq. (\ref{QCD}) and the
complexity length scale \textrm{L}$_{\mathrm{C}}\left(  \alpha\right)  $ in
Eq. (\ref{complexityL}) for two supplementary angles, $\alpha_{1}%
\overset{\text{def}}{=}(1/16)\pi$ and $\alpha_{2}\overset{\text{def}}{=}%
\pi-\alpha_{1}=(15/16)\pi$. A similar calculation scheme applies to any other
pair of supplementary angles $\alpha$ and $\pi-\alpha$, with $0\leq\alpha
\leq\pi$. Clearly, the essential quantities needed to arrive at the estimates
of \textrm{C}$\left(  \alpha\right)  $ and \textrm{L}$_{\mathrm{C}}\left(
\alpha\right)  $ are the accessed volume $\overline{\mathrm{V}}\left(
\alpha\text{, }\omega\right)  $ in Eq. (\ref{avgcomplexity}), the accessible
volume \textrm{V}$_{\max}\left(  \alpha\text{, }\omega\right)  $ in Eq.
(\ref{j4}), and the path length $s\left(  \alpha\right)  $ in Eq.
(\ref{yoyo}). We recall that we set in what follows $\hslash=1$ and
$\omega\overset{\text{def}}{=}E/\hslash=1$. Although we do not use physical
units in an explicit fashion, assume that the usual MKSA unit system is
employed here.

\subsection{Case: $\alpha_{1}=\pi/16$}

For $\alpha_{1}=\pi/16$, the time-dependent volume $V\left(  t\right)  $ in
Eq. (\ref{vt}) becomes%
\begin{equation}
V(t)=\frac{1}{4}\left\vert \varphi\left(  t\right)  \cos\left[  \theta\left(
t\right)  \right]  \right\vert \text{.} \label{bill1}%
\end{equation}
The polar and azimuthal angles in Eq. (\ref{bill1}) are given by,%
\begin{equation}
\theta\left(  t\right)  \overset{\text{def}}{=}2\arctan\left(  \sqrt
{\frac{\frac{1}{2}-\frac{\sin(\frac{\pi}{8})}{2\sqrt{2}}\sin^{2}(t)+\frac
{\cos(\frac{\pi}{16})}{2\sqrt{2}}\sin(2t)}{\frac{1}{2}+\frac{\sin(\frac{\pi
}{8})}{2\sqrt{2}}\sin^{2}(t)-\frac{\cos(\frac{\pi}{16})}{2\sqrt{2}}\sin(2t)}%
}\right)  \text{,} \label{bill2}%
\end{equation}
and,%
\begin{equation}
\varphi\left(  t\right)  \overset{\text{def}}{=}\left\{
\begin{array}
[c]{c}%
\varphi_{1}\left(  t\right)  \overset{\text{def}}{=}\arctan\left(  \frac
{\frac{\sin(\frac{\pi}{16})\sin(t)}{\sqrt{2}}-\frac{\cos(\frac{\pi}{16}%
)\sin(t)}{2}}{\frac{\cos(\frac{\pi}{16})\sin(t)}{2}+\frac{\cos(t)}{\sqrt{2}}%
}\right)  -\arctan\left(  \frac{\frac{\cos(\frac{\pi}{16})\sin(t)}{2}%
+\frac{\sin(\frac{\pi}{16})\sin(t)}{\sqrt{2}}}{\frac{\cos(\frac{\pi}{16}%
)\sin(t)}{2}-\frac{\cos(t)}{\sqrt{2}}}\right)  \text{, for }0\leq t\leq
t_{1}\\
\\
\varphi_{2}\left(  t\right)  \overset{\text{def}}{=}\arctan\left(  \frac
{\frac{\sin(\frac{\pi}{16})\sin(t)}{\sqrt{2}}-\frac{\cos(\frac{\pi}{16}%
)\sin(t)}{2}}{\frac{\cos(\frac{\pi}{16})\sin(t)}{2}+\frac{\cos(t)}{\sqrt{2}}%
}\right)  -\arctan\left(  \frac{\frac{\cos(\frac{\pi}{16})\sin(t)}{2}%
+\frac{\sin(\frac{\pi}{16})\sin(t)}{\sqrt{2}}}{\frac{\cos(\frac{\pi}{16}%
)\sin(t)}{2}-\frac{\cos(t)}{\sqrt{2}}}\right)  +\pi\text{, for }t_{1}\leq
t\leq t_{2}%
\end{array}
\right.  \text{,} \label{bill3}%
\end{equation}
respectively, with $t_{1}\overset{\text{def}}{=}\arctan\left\{  \sqrt
{2}/\left[  \cos(\frac{\pi}{16})\right]  \right\}  \simeq0.9644$ and
$t_{2}\simeq1.3781$ being the time needed to arrive from the initial state
$\left\vert A\right\rangle $ to the final state $\left\vert B\right\rangle $.
Using Eqs. (\ref{bill1}), (\ref{bill2}), and (\ref{bill3}), the accessed
volume $\overline{\mathrm{V}}$ becomes%
\begin{equation}
\overline{\mathrm{V}}=\frac{1}{t_{1}}\int_{0}^{t_{1}}\frac{1}{4}\left\vert
\varphi_{1}\left(  t\right)  \cos\left[  \theta\left(  t\right)  \right]
\right\vert dt+\frac{1}{t_{2}-t_{1}}\int_{t_{1}}^{t_{2}}\frac{1}{4}\left\vert
\varphi_{2}\left(  t\right)  \cos\left[  \theta\left(  t\right)  \right]
\right\vert dt\text{.} \label{bill4}%
\end{equation}
From a numerical integration of the right-hand-side in Eq. (\ref{bill4}), we
get $\overline{\mathrm{V}}=6.\,538\,5\times10^{-2}+8.\,823\,8\times
10^{-2}\simeq0.153\,6$. Next, to get an estimate of \textrm{V}$_{\max}$, we
need to find $\theta_{\min}$, $\theta_{\max}$, $\varphi_{\min}$, and
$\varphi_{\mathrm{\max}}$. Inspecting Eqs. (\ref{bill2}) and (\ref{bill3}), we
obtain $\theta_{\min}=\theta\left(  0\right)  =\pi/2\simeq1.5708$,
$\theta_{\max}=\theta((t_{B}-t_{A})/2)=\theta\left(  1.3781/2\right)
\simeq2.1789$, $\varphi_{\min}=\varphi(0)=0$, and $\varphi_{\max}%
=\varphi(t_{B}-t_{A})=\varphi(1.3781)=\pi/2\simeq1.5708$. Therefore,
\textrm{V}$_{\max}$ is equal to%
\begin{equation}
\mathrm{V}_{\max}=\frac{1}{4}\left(  \int_{\theta_{\min}}^{\theta_{\max}}%
\sin(\theta)d\theta\right)  \left(  \int_{\varphi_{\min}}^{\varphi_{\max}%
}d\varphi\right)  \text{,} \label{bill5}%
\end{equation}
that is, $\mathrm{V}_{\max}\simeq0.2243$. From Table III, we see that
$s=1.9857$ for $\alpha_{1}=\pi/16$. Therefore, using the numerical estimates
of $\overline{\mathrm{V}}$, $\mathrm{V}_{\max}$, and $s$, we finally arrive at
\textrm{C}$\simeq0.3152$ and $\mathrm{L}_{\mathrm{C}}\simeq2.3996$. These are
the values that appear in Tables I and II.

\subsection{Case: $\alpha_{2}=(15/16)\pi$}

For $\alpha_{2}=(15/16)\pi$, the time-dependent volume $V\left(  t\right)  $
in Eq. (\ref{vt}) become%
\begin{equation}
V(t)=\frac{1}{4}\left\vert \varphi\left(  t\right)  \cos\left[  \theta\left(
t\right)  \right]  \right\vert \text{.} \label{bill1b}%
\end{equation}
The polar and azimuthal angles in Eq. (\ref{bill1}) are given by,%
\begin{equation}
\theta\left(  t\right)  \overset{\text{def}}{=}2\arctan\left(  \sqrt
{\frac{\frac{1}{2}-\frac{\sin(\frac{15}{8}\pi)}{2\sqrt{2}}\sin^{2}%
(t)+\frac{\cos(\frac{15}{16}\pi)}{2\sqrt{2}}\sin(2t)}{\frac{1}{2}+\frac
{\sin(\frac{15}{8}\pi)}{2\sqrt{2}}\sin^{2}(t)-\frac{\cos(\frac{15}{16}\pi
)}{2\sqrt{2}}\sin(2t)}}\right)  \text{,} \label{bill6}%
\end{equation}
and,%
\begin{equation}
\varphi\left(  t\right)  \overset{\text{def}}{=}\left\{
\begin{array}
[c]{c}%
\varphi_{1}\left(  t\right)  \overset{\text{def}}{=}\arctan\left(  \frac
{\frac{\sin(\frac{15}{16}\pi)\sin(t)}{\sqrt{2}}-\frac{\cos(\frac{15}{16}%
\pi)\sin(t)}{2}}{\frac{\cos(\frac{15}{16}\pi)\sin(t)}{2}+\frac{\cos(t)}%
{\sqrt{2}}}\right)  -\arctan\left(  \frac{\frac{\cos(\frac{15}{16}\pi)\sin
(t)}{2}+\frac{\sin(\frac{15}{16}\pi)\sin(t)}{\sqrt{2}}}{\frac{\cos(\frac
{15}{16}\pi)\sin(t)}{2}-\frac{\cos(t)}{\sqrt{2}}}\right)  \text{, for }0\leq
t\leq t_{1}\\
\\
\varphi_{2}\left(  t\right)  \overset{\text{def}}{=}\arctan\left(  \frac
{\frac{\sin(\frac{15}{16}\pi)\sin(t)}{\sqrt{2}}-\frac{\cos(\frac{15}{16}%
\pi)\sin(t)}{2}}{\frac{\cos(\frac{15}{16}\pi)\sin(t)}{2}+\frac{\cos(t)}%
{\sqrt{2}}}\right)  -\arctan\left(  \frac{\frac{\cos(\frac{15}{16}\pi)\sin
(t)}{2}+\frac{\sin(\frac{15}{16}\pi)\sin(t)}{\sqrt{2}}}{\frac{\cos(\frac
{15}{16}\pi)\sin(t)}{2}-\frac{\cos(t)}{\sqrt{2}}}\right)  +\pi\text{, for
}t_{1}\leq t\leq t_{2}%
\end{array}
\right.  \text{,} \label{bill7}%
\end{equation}
respectively, with $t_{1}\overset{\text{def}}{=}\arctan\left\{  \sqrt
{2}/\left[  \cos(\frac{\pi}{16})\right]  \right\}  \simeq0.9644$ and
$t_{2}\simeq1.3781$ being the time needed to arrive from the initial state
$\left\vert A\right\rangle $ to the final state $\left\vert B\right\rangle $.
To find $t_{1}$, we used the fact that $\cos\left(  \pi/16\right)
=-\cos\left[  (15/16)\pi\right]  $. Using Eqs. (\ref{bill1b}), (\ref{bill6}),
and (\ref{bill7}), the accessed volume $\overline{\mathrm{V}}$ reduces to%
\begin{equation}
\overline{\mathrm{V}}=\frac{1}{t_{1}}\int_{0}^{t_{1}}\frac{1}{4}\left\vert
\varphi_{1}\left(  t\right)  \cos\left[  \theta\left(  t\right)  \right]
\right\vert dt+\frac{1}{t_{2}-t_{1}}\int_{t_{1}}^{t_{2}}\frac{1}{4}\left\vert
\varphi_{2}\left(  t\right)  \cos\left[  \theta\left(  t\right)  \right]
\right\vert dt\text{.} \label{bill4b}%
\end{equation}
From a numerical integration of the right-hand-side in Eq. (\ref{bill4b}), we
get $\overline{\mathrm{V}}=6.\,538\,5\times10^{-2}+8.\,823\,8\times
10^{-2}\simeq0.153\,6$. Next, to get an estimate of \textrm{V}$_{\max}$, we
need to find $\theta_{\min}$, $\theta_{\max}$, $\varphi_{\min}$, and
$\varphi_{\mathrm{\max}}$. Inspecting Eqs. (\ref{bill6}) and (\ref{bill7}), we
obtain $\theta_{\min}=\theta\left(  (t_{B}-t_{A})/2\right)  =\theta\left(
1.3781/2\right)  \simeq0.9627$, $\theta_{\max}=\theta(0)=\pi/2\simeq1,5708$,
$\varphi_{\min}=\varphi\left(  0\right)  =0$, and $\varphi_{\max}%
=\varphi\left(  t_{B}-t_{A}\right)  =\varphi\left(  1.3781\right)  =$
$\pi/2\simeq1.5708$. Therefore, \textrm{V}$_{\max}$ equals%
\begin{equation}
\mathrm{V}_{\max}=\frac{1}{4}\left(  \int_{\theta_{\min}}^{\theta_{\max}}%
\sin(\theta)d\theta\right)  \left(  \int_{\varphi_{\min}}^{\varphi_{\max}%
}d\varphi\right)  \text{,}%
\end{equation}
that is, $\mathrm{V}_{\max}\simeq0.2243$. From Table III, we see that
$s=1.9857$ for $\alpha_{2}=(15/16)\pi$. Therefore, using the numerical
estimates of $\overline{\mathrm{V}}$, $\mathrm{V}_{\max}$, and $s$, we finally
arrive at \textrm{C}$\simeq0.3152$ and $\mathrm{L}_{\mathrm{C}}\simeq
2.3996$.\ These are the values that are displayed in Tables I and
II.\begin{table}[t]
\centering
\begin{tabular}
[c]{c|c|c|c}\hline\hline
$\alpha$ & $t\left(  \alpha\right)  $ & $s\left(  \alpha\right)  $ &
$\pi-\alpha$\\\hline
$0$ & $1.5708$ & $2.2214$ & $\pi$\\\hline
$\frac{1}{16}\pi$ & $1.3781$ & $1.9857$ & $\frac{15}{16}\pi$\\\hline
$\frac{1}{8}\pi$ & $1.2053$ & $1.8251$ & $\frac{7}{8}\pi$\\\hline
$\frac{3}{16}\pi$ & $1.0637$ & $1.7208$ & $\frac{13}{16}\pi$\\\hline
$\frac{1}{4}\pi$ & $0.9553$ & $1.6547$ & $\frac{3}{4}\pi$\\\hline
$\frac{5}{16}\pi$ & $0.8772$ & $1.6133$ & $\frac{11}{16}\pi$\\\hline
$\frac{3}{8}\pi$ & $0.8249$ & $1.5883$ & $\frac{5}{8}\pi$\\\hline
$\frac{7}{16}\pi$ & $0.7951$ & $1.5750$ & $\frac{9}{16}\pi$\\\hline
$\frac{1}{2}\pi$ & $0.7854$ & $1.5708$ & $\frac{1}{2}\pi$\\\hline
\end{tabular}
\caption{Numerical estimates for the evolution time $t\left(  \alpha\right)  $
and the length of the path $s\left(  \alpha\right)  $, with $0\leq\alpha
\leq\pi$. The interval $\left[  0\text{, }\pi\right]  $ was partitioned into
sixteen sub-intervals of equal size $\pi/16$. We set $\hslash=1$ and
$\omega=1$ in the calculations.}%
\end{table}
\end{document}